\documentclass[a4paper,10pt]{article}
\usepackage{amsmath,amsfonts,amsthm}
\usepackage{graphicx}
\usepackage{cite}
\usepackage{hyperref}
\usepackage[title]{appendix}
\newcommand{\dd}{\textrm{d}}

\begin{document}
\title{\bf The expectation value of the number of loops\\ and the left-passage probability in the \\double-dimer model}
\author{Nahid Ghodratipour\footnote{ghodratipour\_n@physics.sharif.edu}\quad and\quad Shahin Rouhani\footnote{srouhani@sharif.ir} \\
\small Department of Physics, Sharif University of Technology, P.O. Box 11155-9161, Tehran, Iran}
\maketitle

%=========================================================================%
\begin{abstract}
We study various statistical properties of the double-dimer model, a generalization of the dimer model, on rectangular domains of the square lattice. 
We take advantage of the Grassmannian representation of the dimer model, first to calculate the probability distribution of the number of nontrivial loops around a cylinder, which is consistent with the previously known result, and then to calculate the expectation value of the number of loops surrounding two faces and the left-passage probability, both in the discrete and the continuum cases. We also briefly explain the calculation of some related observables.
As a by-product, we obtain the partition function of the dimer model in the presence of two and four monomers, and a single monomer on the boundary.
\end{abstract}

%=========================================================================%
\section{Introduction}
\label{introduction}
Some recent activities have focused on mathematical theories of critical phenomena, namely Schramm-Loewner Evolution (SLE) and Conformal Loop Ensemble (CLE) \cite{schramm, schlawwer, rohdschram, smirtoward, katori, werself, sheffield,burgcamialis}. SLEs are random planar curves which are characterized by certain conformal invariance property, following a Loewner evolution driven by a Brownian motion with positive diffusivity $\kappa$ \cite{wernerrev,kager,rouh}. It is a powerful tool in the study of the scaling limits of two-dimensional critical models, which in several cases such as the loop-erased random walk and uniform spanning tree, percolation and the Ising model on specific lattices, the convergence has been established rigorously. CLEs are ensembles of countably random collection of planar conformally invariant non-crossing and non-self-crossing loops in simply connected domains of the complex plane, indexed by a parameter $\kappa$ in $(8/3,8]$ and constructed by branching variants of SLEs \cite{sheffield}. Both ideas are interesting to physicists because they are related to interfaces in critical phenomena, where early attentions to SLE were mainly motivated by Smirnov's proof of Cardy's formula, which in turn leads to the convergence of the percolation exploration path to the chordal SLE$_6$. Further the scaling limit of the Ising model was proved to be SLE$_3$ or SLE$_{16/3}$, determined by the representation \cite{smirnov,chelduminhongkempsmir}. Alternatively, it is related to a conformally invariant loop model at criticality hence a CLE, with either $\kappa=3$ or $\kappa=16/3$.

Loop models are among most important and richest classes of two-dimensional statistical lattice models and indeed, many statistical mechanical models are best understood in terms of random paths, open or closed, at criticality \cite{saleur,duplantier,vanderzande}. In addition to providing geometrical explanations and tools, it employs many methods to study scaling limits, including Coulomb gas formalism \cite{nienhuis}, Conformal Field Theory (CFT) \cite{henkel} and more recently, rigorous theories such as SLE and CLE \cite{cardy,sheffwer}. In all these theories conformal invariance plays a crucial role, enables them to determine critical exponents and other associated universal properties of two-dimensional critical systems such as correlation functions exactly. SLEs (CLEs) are believed to be the scaling limits of various random loop models from statistical physics, including the O($n$) models. These are a large class of critical systems in physics, conjectured to be related to SLE$_\kappa$ and CLE$_\kappa$ concisely expressed by the formula $n=-2\cos(4\pi/\kappa)$. The parameter $\kappa$ is so in significant relation to the central charge $c$ in CFT and as a fact, SLEs (CLEs) with $\kappa\leq 4$ are believed to be the minimal models of CFT with $c$ in $[0,1]$ \cite{bb,friwer1,friwer2}. One prototypical model where such issues are concerned is the two-dimensional Gaussian Free Field (GFF).

The two-dimensional GFF is a central object both in physics and mathematics, related to the Coulomb gas, which falls into the same universality class as other statistical mechanical models such as the XY-model and the two-dimensional sine-Gordon model \cite{nienhuis,schsheff,kenyonintro}. It is the mathematical model of two-dimensional bosonic CFT, and is a two-dimensional analog of the Brownian motion \cite{sheff}. The notions associated with the Brownian motion, such as the Gaussian measure, the Laplacian, the Green's function and consequently conformal invariance, are also substantial in GFF \cite{kenyonintro,kendirac}, and many stochastic differential equations in two dimensions depend on it as those in one dimension on the Brownian motion \cite{karparzh,borferr}. We can practically think of GFF as a natural model of random height functions though it is actually a generalized function or a distribution (in sense of Schwartz \cite{sheff}). It yields a dual interpretation of many loop models, especially the O($n$) models, which in SLE/CLE formalism, is very natural in the case of $\kappa=4$ (equivalently $n=2$); the zero level line of the Gaussian free field (with appropriate boundary conditions) is a chordal SLE$_4$ and the collection of level loops of it corresponds to CLE$_4$ \cite{dubedat,mengwu}.

Besides, there is a family of conformally invariant loop models in two dimensions, the Brownian Loop Soups (BLSs), which is deeply related to SLE and CLE \cite{werner,sheffwer}. Each BLS is a Poisson point process of Brownian loops \cite{lawwerner}, with intensity parameter related to $\kappa$ via the equation $c=(3\kappa-8)(6-\kappa)/2\kappa$ \cite{werner}, indicating that for low enough intensities the outer boundaries of clusters of Brownian loops are distributed like CLEs for the spectral parameter between 8/3 and 4. This implies that the intensity parameter is the central charge of the relevant CFT for $c$ in $[0,1]$. Motivated by the relation between BLS, SLE and CLE, a connection between BLS and CFT was discovered analyzing particular expectation values, such as correlation functions of the layering and the winding number operators \cite{camia}. The critical BLS ($c=1$) is closely related to the Gaussian Free Field.

All these relations allow or rather require to exploit achievements of mathematical theories of critical phenomena in physics. In this regard, it is quite standard to find and compute observables in critical lattice models, evidencing and supporting relations to SLE and CLE, and/or determining $\kappa$. This is what we have done in this paper, in the case of the double-dimer model. Such efforts are prevalent in statistical physics literature. 

The study of relevant SLEs has been performed over the last decade with some degree of success. However, not much work has been done on CLEs by physicists. This is perhaps due to the mathematical tractability of CLEs. To be specific, many two-dimensional physical models result in random surfaces whose contour lines are CLEs, yet in physics literature they are inevitably cut in half and treated as SLEs \cite{darya,wo3,kobayashi}. 

On the other hand, there are limitations on describing a critical model through SLE. The existence of an SLE curve a priori depends on imposing certain boundary conditions, and many observables, potentially defined on cluster boundaries, could not be supported by SLE. Finding tractable methods of tackling CLE is therefore of importance to the physics community. One way of analyzing CLE is via the double-dimer model which is a straightforward generalization of the dimer model.

The dimer model is the study of canonical measures, i.e. Gibbs measures, on the set of dimer coverings of a graph \cite{kenyonintro}. A dimer covering or perfect matching of a graph is a collection of occupied edges (dimers) of the graph, such that every vertex is covered exactly once. Dimers can have different fugacities (Boltzmann weights), and the weight of each dimer cover is defined to be the product of all dimer weights in the configuration. Generally, interesting graphs are periodic planar graphs such as the square lattice or the honeycomb lattice.

A double-dimer configuration is a union of two dimer covers, or equivalently a set of even-length simple loops and doubled edges, with the property that every vertex is the endpoint of exactly two edges (which may be doubled) \cite{kenyondouble}. This immediately leads to a loop model related to CLE; it is interesting to study the properties of the distribution of long loops, especially the conformal invariance of the scaling limit of the "uniform" double-dimer model, and its relation to CLE$_4$/SLE$_4$. The partition function of the double-dimer model is simply the multiplication of partition functions of the two contributing dimer models, giving a proportionally weight of $2^k$ to each configuration with k loops in the uniform case. The double-dimer model offers a tool to study the loop model hence CLE.

Precisely in the case of dimers on a bipartite graph, each configuration can be mapped to a height function, a particular discretization of the GFF. Strongly evidenced by this fact, which proved by Kenyon for Temperleyan approximations of $\mathbb{Z}^2$  \cite{kenyondominoconf, kenyondomino}, he also predicted the convergence of double-dimer interfaces and loop ensembles to SLE$_4$ and CLE$_4$, respectively. Despite pioneer works afterward by Kenyon \cite{kenyondouble} and Dubedat \cite{dubedatdeformation} on the convergence of topological observables defined on double-dimer loops, and the most recent complementing results in \cite{basokchelkak}, this prediction has not fully proven yet. Not to mention the instance of an underlying nonbipartite graph, where any precise statement in this direction is still unavailable.\footnote{We however refer the interested reader to \cite{dimertori}, for the very suggestive results on this issue.}

The dimer model arose initially in physics studying the adsorption of diatomic molecules on a surface \cite{fowlerrush}, later an abstraction and an exact formal solution for planar graphs was proposed by Kasteleyn \cite{kasteleyn}, and independently by Fisher and Temperley \cite{temperleyfisher}. In the case of the square lattice, the exact number of domino-tilings of an $M\times N$ (both even) surface by $2\times 1$ dominos is given by:
\begin{align}
\mathcal{Q}_0=\prod_{p=1}^{\frac{M}{2}}\prod_{q=1}^{\frac{N}{2}}\left[4\cos^2\frac{\pi p}{M+1}+4\cos^2\frac{\pi q}{N+1}\right]
\label{uniformpartition}
\end{align}
which is also the partition function of the dimer model in the uniform case (all dimer fugacities are 1). Since then, corresponding computations have been done for some other lattices (see \cite{wu} for a review of various references).

The dimer model is of interest in its own right. While not much practical as a realistic model, the dimer model has inherent interest as an exactly solvable model whose distinguished types of phase transitions can be studied analytically. It has close relation with random matrices and determinantal processes \cite{cohnelkprop,kenyonlocal,cohnkenprop,boutillier} and to begin with, the computation of the partition function can be accomplished using determinants \cite{kasteleyn,temperleyfisher}. Apart from its connection to many contexts in mathematics, it can be translated to several other statistical mechanical models. There is a correspondence between the two-dimensional Ising model on a graph and the dimer model on a decorated version of this graph \cite{kasteleynising,fisher}, makes dimer techniques a supplementary powerful tool to study the Ising model. It also has relations with vertex models \cite{kenmilshefwil}. As there are few efficient methods to study the latter, these relations are highly valuable \cite{baxter,kenyonintro}.

The fermionic nature of the dimer model, already implied by its relation to the Ising model, is an important feature manifest in Kasteleyn's Pfaffian solution of the dimer problem. Identically, it exhibits in reformulation of the dimer model in terms of Grassmannian anticommuting variables, which leads to an exact solution of the close-packed dimer problem on some lattices, especially the square lattice \cite{hp}. This method works even when there are some monomers present, results in a more complex exact solution in the case of the boundary monomer-dimer problem, and a formal expression of the partition function in the presence of bulk monomers \cite{allegra,jerrum}. 

The central idea of this paper is to use the fermionic representation of the dimer model to give us a tool for direct study of the CLEs and in fact, to utilize this connection for the exact calculation of certain invariants and correlators.

This paper is organized as follows. In section \ref{grassmann}, we review the fermionic representation of the dimer model on the square lattice. We introduce our method, strongly based on \cite{hp}, and perform the calculation of the partition function of the dimer model on a rectangular domain with free and cylindrical boundary conditions. Although the combinatorics we use is somewhat elementary, we give it in the appendix \ref{calculation}, since similar expressions appear over and over in the sequence. A connection to Chebyshev polynomials is also justified here. In section \ref{annulus}, we look at loops around a cylinder and find the distribution of the number of nontrivial loops, which is identical to the previously known result \cite{kenyondouble}. By approximating the result for a very long cylinder, a connection to the Coulomb gas is established. In section \ref{loopssurrounding}, we calculate the expectation value of the number of loops surrounding two points, which in the scaling limit, is in accordance with the prediction of the relation between double-dimer loop ensembles, CLE$_4$ and the level lines of the GFF at certain heights. We also compute the one-point function and comment on the calculation of some related observables. Otherwise mentioned, all loops are unoriented in this paper. In section \ref{monomer}, we revisit the boundary monomer-dimer problem and calculate the partition function in the presence of two and four monomers, and a single monomer on the boundary. We can particularly check the latter to be in agreement with the result in \cite{allegra}. In section \ref{leftpassage}, the left-passage probability is addressed and we find the approximation of the result for a very large lattice consistent with what is expected for the chordal SLE$_4$ on a rectangular domain. Some less relevant, but still useful formulas have been moved to the appendix \ref{useful}. We also obtain a hitherto unknown series identity in the appendix \ref{seriesformula}. 
%=============================================================================%
\section{Grassmannian representation of the dimer model}
\label{grassmann}
For any dimer model on a finite graph, there is an adjacency matrix which has indices representing vertices of the graph, and entries equal the fugacities of bonds and zero otherwise. The partition function of the dimer model is then the hafnian of the adjacency matrix, which in the particular case of a bipartite graph is the square root of the permanent. However, even the permanent of a $(0,1)$-matrix, which appears to be a less computationally complicated object than the hafnian, is not tractable in general \cite{valiant}. A theorem due to Kasteleyn \cite{kasteleyn} guarantees that for any planar graph, there is an associated signed adjacency matrix, a Kasteleyn matrix, which the partition function of the dimer model is the Pfaffian (the square root of the determinant) of this matrix. The partition function of the double-dimer model is then the determinant of the relevant Kasteleyn matrix. Here we study the uniform double-dimer model (based on two uniform dimer model) on finite sublattices of the square grid, by a priori non-combinatorial approach through Grassmannian representation of the dimer model.

For pedagogical reasons, we first compute the partition function of the uniform dimer model (equiprobable dimer covers) when the square lattice is embedded on the cylinder, i.e. the boundary conditions are periodic and free along the horizontal and vertical directions respectively
\footnote{We should clarify that by the free boundary conditions we mean that, at the current situation, all vertical bonds just below the bottom side and just above the top side of the rectangle are removed, so that on these sides, it is diconnected from the surroundings. It corresponds to a situation where dimers are not allowed to protrude accross the boundary.} 
. For a lattice of size $M\times N$, $M$ and/or $N$ even, it can be written as
\begin{align}
   \mathcal{Q}_0=\int \prod_{n=1}^{N}\prod_{m=1}^{M} \dd\eta_{m,n} (1+\eta_{m,n}\eta_{m+1,n}) (1+\eta_{m,n}\eta_{m,n+1})
  \label{nilpartition}
\end{align}
with the periodic/free boundary conditions: $\eta_{M+1,n}=\eta_{1,n}$, $\eta_{m,N+1}=0$, where $\eta_{m,n}$ are commuting nilpotent variables attached to every site, satisfying $\eta_{m,n}^2=0$, to prevent double occupancy of a site by two dimers, and $\int\dd\eta_{m,n}\eta_{m,n}=1$ and $\int\dd\eta_{m,n}=0$, to ensure that each site is met by a dimer. The zero index of the partition function indicates the non-existence of defects (monomers) in the system, i.e. we deal with the pure close-packed dimer problem. 

The representation \eqref{nilpartition} of the partition function is identical to the Haffnian of the adjacency matrix of the underlying square lattice. Confirmed with Kasteleyn's theorem, we have to replace \eqref{nilpartition} with a representation identical to the Pfaffian of some related matrix. Following Hayn and Plechko \cite{hp} closely, we can introduce a set of completely anticommuting Grassmann variables 
$\{a_{m,n}, \bar{a}_{m,n}\}$
 and
$\{b_{m,n}, \bar{b}_{m,n}\}$
(so with the same calculus as $\{\eta_{m,n}\}$)
for horizontal and vertical bonds respectively, such that
\begin{align}
  &1+\eta_{m,n}\eta_{m+1,n}=\int \dd {\bar{a}}_{m,n}\dd a_{m,n} \ \mathrm{e}^{a_{m,n}{\bar{a}}_{m,n}} (1+a_{m,n}\eta_{m,n})(1+{\bar{a}}  _{m,n}\eta_{m+1,n})\nonumber\\
 & 1+\eta_{m,n}\eta_{m,n+1}=\int \dd {\bar{b}}_{m,n}\dd b_{m,n}\ \mathrm{e}^{b_{m,n}{\bar{b}}_{m,n}} (1+b_{m,n}\eta_{m,n})(1+{\bar{b}}_{m,n}\eta_{m,n+1}).
\label{decouple}
\end{align}
We further introduce the notation 
$\{A_{m,n}, \bar{A}_{m,n}, B_{m,n}, \bar{B}_{m,n}\}$,
\begin{align}
  &A_{m,n}=1+a_{m,n}\eta_{m,n},\hspace{1.2cm}\bar{A}_{m+1,n}=1+{\bar{a}}  _{m,n}\eta_{m+1,n}\nonumber\\
  &B_{m,n}=1+b_{m,n}\eta_{m,n},\hspace{1.2cm}\bar{B}_{m,n+1}=1+{\bar{b}}_{m,n}\eta_{m,n+1}\nonumber
\end{align}
which are simply non-commuting Grassmann factors made use of for brevity. Thanks to the mirror-ordering method \cite{pIsing}, the partition function can then be written as
\begin{align}
\mathcal{Q}_0= \mathrm{Sp}_{\{a,\bar{a},b,\bar{b},\eta\}} \prod_{n=1}^{N} \left(\prod_{m=1}^{M}\overset{m}{\overleftarrow{\bar{B}_{m,n}}} \prod_{m=1}^{M} \bar{A}_{m,n}\overset{m}{\overrightarrow{B_{m,n}}}A_{m,n}\right),
\label{mirror}
\end{align}
where $\mathrm{Sp}_{\{.\}}$ stands for the Gaussian averaging $\int\dd\bar{a}\hspace{.04cm}\dd a\hspace{.04cm}\dd\bar{b}\hspace{.04cm}\dd b\hspace{.1cm}e^{a\bar{a}+b\bar{b}}(...)$, as well as the integration over nilpotent variables $\{\eta\}$, and arrows indicate the direction of increasing in the index "m".
This way, the individual $\eta_{mn}$ can be isolated and integrated to yield a purely Grassmannian representation of the partition function of the dimer model: 
\begin{align}
  \mathcal{Q}_0=\mathrm{Sp}_{\{a,\bar{a},b,\bar{b}\}} \prod_{m,n} L_{m,n},\hspace{0.8cm}L_{m,n}=a_{m,n}+b_{m,n}+\bar{a}_{m-1,n}+(-1)^{m+1}\bar{b}_{m,n-1}.\nonumber
\end{align}
where the factor $(-1)^{m+1}$ arises as the result of translating anticommuting variables $L_{m,n}$ through Grassmann factors $\bar{B}_{m,n}$, from meeting point to the left. With the aid of additional Grassmann variables $c_{m,n}$ defined by the identity $L_{m,n}=\int\dd c_{m,n}\exp (c_{m,n}L_{m,n})$, the final expression is:
\begin{align}
  \mathcal{Q}_0=\int\prod_{n=1}^{N}\prod_{m=1}^{M}\overset{m}{\overrightarrow{\dd c_{m,n}}}\  \exp\  
  \left\{\sum_{m=1}^{M}\sum_{n=1}^{N} [c_{m+1,n}c_{m,n}+(-1)^{m+1}c_{m,n+1}c_{m,n}]\right\}
 \label{grasspartition}
\end{align}
where we call $\mathcal{S}_0$ the fermionic action of the close-packed dimer model,
\begin{align}
-\mathcal{S}_0=\sum_{m=1}^{M}\sum_{n=1}^{N} [c_{m+1,n}c_{m,n}+(-1)^{m+1}c_{m,n+1}c_{m,n}]
\label{actionpure}
\end{align}
in parallel to the path integral formalism in field theory. 

Without loss of generality, we assume that $\frac{M}{2}$ is an even integer. This imposes aperiodic boundary conditions for fermions (Grassmann variables) in horizontal direction,
$c_{M+1,n}=-c_{1,n}$,
 because of the transposition of the products of the boundary Grassmann factors. Combinatorially, it is identical to compensating the minus sign due to the even cyclic permutation around the cylinder in the expansion of the determinant of Kasteleyn's matrix. As a result, we have to use half-integer momenta in Fourier substitution (in horizontal direction) for fermionic variables. Remember free boundary conditions in vertical direction,
$c_{m,N+1}=0$,
 we can pass into momentum space through the transformation \cite{hp}
\begin{align}
 c_{m,n}=\sqrt{\frac{2}{M(N+1)}} i^{n}\sum_{p=1}^{M}\sum_{q=1}^{N}c_{p,q}\ \mathrm{e}^{i\frac{\pi(2p-1)m}{M}}\sin\frac{\pi qn}{N+1}
\label{fullfourier}
\end{align}
Using the rules \eqref{exp} and \eqref{sin}, we can sum up the terms in \eqref{actionpure} and obtain the ultimate coupling between momentum modes
\begin{align*}
-\mathcal{S}_0=\sum_{p,p'}\sum_{q,q'} c_{p,q}c_{p',q'}[-\mathrm{e}^\frac{i\pi(2p+1)}{M}\delta_{q+q',N+1}\delta_{p+p'-1,M}+i\cos\frac{\pi q}{N+1}\delta_{q+q',N+1}(\delta_{p+p'-1,\frac{M}{2}}+\delta_{p+p'-1,\frac{3M}{2}})]
\end{align*} 
Therefore the partition function \eqref{grasspartition} factorizes into the product of the Pfaffians of similar antisymmetric matrices
\begin{align}
\left[ {\begin{array}{cc} 0&A \\ -A&0 \end{array}} \right]
\label{matrix}
\end{align}
where, with a little abuse of notation $-q\equiv N+1-q$, the matrix $A$
\begin{align}
A=\left[ {\begin{array}{cccc} 0&2i\cos\frac{\pi q}{N+1}&0&-2i\sin\frac{\pi(2p-1)}{M} \\ -2i\cos\frac{\pi q}{N+1}&0&-2i\sin\frac{\pi(2p-1)}{M}&0 \\ 0&2i\sin\frac{\pi(2p-1)}{M}&0&2i\cos\frac{\pi q}{N+1} \\ 2i\sin\frac{\pi(2p-1)}{M}&0&-2i\cos\frac{\pi q}{N+1}&0 \end{array}} \right]
\label{matrixA}
\end{align}
indicates the coupling coefficients of the vector 
$(c_{p,q},c_{\frac{M}{2}-p,q},c_{\frac{M}{2}-1+p,q},c_{M-p,q})^t$
with the vector
$(c_{p,-q},\linebreak c_{\frac{M}{2}-p,-q},c_{\frac{M}{2}-1+p,-q},c_{M-p,-q})$. The factor "2" in the entries in \eqref{matrixA} shows the contribution of first and second half-section of $1\leq p\leq M$ and $1\leq q\leq N$, altogether.
For $N$ even, these lead to the final evaluation of the partition function
\begin{align}
\mathcal{Q}_0^{cyl}=\prod_{p=1}^\frac{M}{2}\prod_{q=1}^\frac{N}{2}\left[4\cos^2\frac{\pi q}{N+1}+4\sin^2\frac{\pi(2p-1)}{M}\right],\hspace{.6cm}(N\hspace{.1cm}\text{even})
\label{cylindereven}
\end{align}
For $N$ odd, the only difference is that there is a single uncoupled mode due to $q=\frac{N+1}{2}$, so the result is
\begin{align}
\mathcal{Q}_0^{cyl}=2\prod_{p=1}^\frac{M}{2}\prod_{q=1}^\frac{N-1}{2}\left[4\cos^2\frac{\pi q}{N+1}+4\sin^2\frac{\pi(2p-1)}{M}\right],\hspace{.6cm}(N\hspace{.1cm}\text{odd})
\label{cylinderodd}
\end{align}
in agreement with the results in \cite{kasteleyn,temperleyfisher}.

The results \eqref{cylindereven} and \eqref{cylinderodd} can also be obtained by a little different method. Indeed for our further analysis of the model, it is more convenient if we do Fourier transformation only for one of the indices and use combinatorics in compensation.

\begin{figure}[h]
	\includegraphics[width=.8\textwidth]{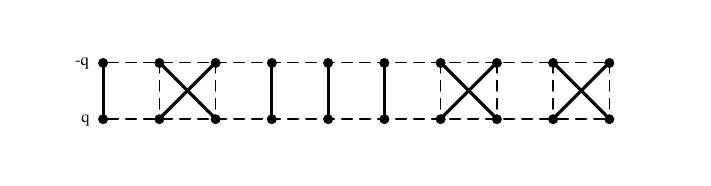}
	\centering
	\caption{An instance of coupled Fourier modes $q$ and $-q$, for a square lattice with $M=10$. Each cross-form has weight one, and each vertical "dimer" has weight $2\cos\frac{\pi q}{N+1}$.}
	\label{fig1}
\end{figure}

Back to the Grassmannian partition function \eqref{grasspartition}, we do Fourier substitution in vertical direction, so \eqref{fullfourier} is replaced by
\begin{align}
 c_{m,n}=\sqrt{\frac{2}{N+1}} i^{n}\sum_{q=1}^{N}c_{m,q}\sin\frac{\pi qn}{N+1}
\label{partfourier}
\end{align}
An overall minus sign or the factor $i$, that may occur in the calculation of partition functions in various situations, does not matter since they are always absorbed by the Jacobian determinant which is trivial here. So we are only concerned with the absolute value of the final results though we have to be careful about the signs of the terms with respect to each other. Applying \eqref{partfourier} the action \eqref{actionpure} becomes
\begin{align}
-\mathcal{S}_0=\sum_{q=1}^N\left\{\sum_{m}\left[-c_{m+1,q}c_{m,-q}-i(-1)^{m+1}\cos\frac{\pi q}{N+1}c_{m,q}c_{m,-q}\right]\right\}
\label{stripaction}
\end{align}
Equation \eqref{stripaction} represents the conversion of the dimer model on the square lattice into some similar model on decoupled strips, each represents the coupling between two Fourier modes $q$ and $-q$. These strips are no longer subgraphs of the square lattice and in fact, they are non-planar as an effect of Fourier transformation (Figure \ref{fig1}). On a given strip, every vertical dimer (with a formal use of the word) has the fugacity, or rather weight of 
$2\cos\frac{\pi q}{N+1}$, again because of the contribution of two half-section of $q$-modes, 
and the contribution of every cross-form leads to the factor 1. We need not worry about $ i $ and the minus signs at the moment, as they would always be absorbed to make every term in the expansion of the partition function have the same sign. For example, we can notice that on every strip, each vertical dimer and its closest next one have opposite parity in position. So, multiplying the coefficient of each vertical dimer with that of the next one yield to a plus sign. For $M$ even, there is an integer number of these couples in every configuration on a strip, and each term is then positive. 

\begin{figure}
	\centering
	\includegraphics[width=.85\textwidth]{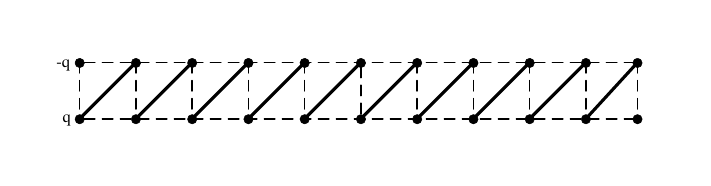}
	\caption{An additional admissible state on account of the cylindrical boundary conditions. The first and last vertical lines are identified.}
	\label{fig2}
\end{figure}

In the absence of block diagonalization of the corresponding Kasteleyn's matrix, we use combinatorics to compute the number of dimer covers on a  cylindrical strip (see the Appendix \ref{calculation}). For $N$ even, the partition function is
\begin{align}
\mathcal{Q}_0^{cyl}=\prod_{q=1}^\frac{N}{2}\left[\left\{\sum_{p=0}^\frac{M}{2}\left[\left( \begin{array}{c} M-p \\ p \end{array} \right)+\left( \begin{array}{c} M-p-1 \\ p-1 \end{array} \right)\right](2\cos\frac{\pi q}{N+1})^{M-2p}\right\} + 2\right],\hspace{.6cm}(N\hspace{.1cm}\text{even})
\label{cylinderpartitioneven}
\end{align}
included the two possibilities of a chain of tilted dimers due to the cylindrical boundary conditions in horizontal direction (Figure \ref{fig2}). For $N$ odd, the single mode $q=\frac{N+1}{2}$ is represented by a one-dimensional dimer model on a line, which has a configuration space with only two members. The partition function becomes
\begin{align}
\mathcal{Q}_0^{cyl}=2\prod_{q=1}^\frac{N-1}{2}\left[\left\{\sum_{p=0}^\frac{M}{2}\left[\left( \begin{array}{c} M-p \\ p \end{array} \right)+\left( \begin{array}{c} M-p-1 \\ p-1 \end{array} \right)\right](2\cos\frac{\pi q}{N+1})^{M-2p}\right\} + 2\right],\hspace{.6cm}(N\hspace{.1cm}\text{odd})
\label{cylinderpartitionodd}
\end{align}

Similarly, we can calculate the partition function of the dimer model with the free boundary conditions, $c_{M+1,n}=0,\hspace{.1cm} c_{m,N+1}=0$. It is
\begin{align}
\mathcal{Q}_0=\prod_{q=1}^\frac{N}{2}\left[\sum_{p=0}^\frac{M}{2}\left( \begin{array}{c} M-p \\ p \end{array} \right)(2\cos\frac{\pi q}{N+1})^{M-2p}\right]
\label{freepartitione}
\end{align}
for $N$ even, and
\begin{align}
\mathcal{Q}_0=\prod_{q=1}^\frac{N-1}{2}\left[\sum_{p=0}^\frac{M}{2}\left( \begin{array}{c} M-p \\ p \end{array} \right)(2\cos\frac{\pi q}{N+1})^{M-2p}\right]
\label{freepartitiono}
\end{align}
for $N$ odd.

We may note that \eqref{cylinderpartitioneven}-\eqref{freepartitiono} are associated with the Chebyshev polynomials of the second kind. We can also check that the partition functions \eqref{cylinderpartitioneven}-\eqref{freepartitiono} are identical to \eqref{cylindereven}, \eqref{cylinderodd} and \eqref{uniformpartition}, with the aid of the identities \eqref{eqkasteleyn} and \eqref{combinatorialidentity} (see the Appendix \ref{calculation}).

In the coming sections, we will use this approach to compute some observables defined on specific collections of double-dimer loops. These observables are supposed to make a strong connection between the scaling limit of the double-dimer model and the SLE and CLE with $\kappa=4$.
%=====================================================================================%
\section{Loops around the cylinder}
\label{annulus}
We can find the distribution of the number of nontrivial loops in the uniform double-dimer model on a cylinder, through the previous semi-combinatorial approach. 
\begin{figure}[h]
	\includegraphics[width=.8\textwidth]{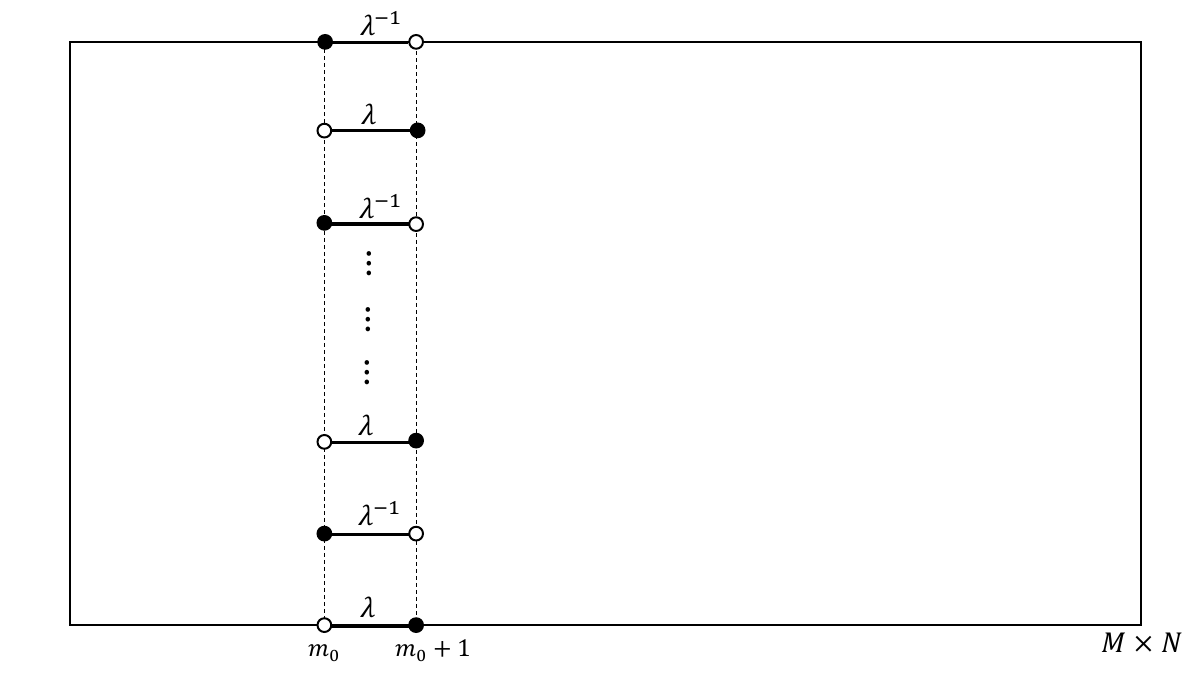}
	\centering
	\caption{The arrangement of fugacities in one of the dimer models. Other bonds have fugacity 1.}
	\label{fig3}
\end{figure}
This case is specially simple because the Kasteleyn matrix can be diagonalized. We arrange some auxiliary fugacities $\lambda$ and $\lambda^{-1}$ in the two contributing dimer models, so that two conditions are satisfied; (1) each nontrivial loop (around the annulus) has to contribute exactly one $\lambda$ or one $\lambda^{-1}$, and (2) trivial loops and loops with zero area (doubled edges) have to contribute the factor $1$, in each term in the expansion of partition function of the double-dimer model. To this end, in one dimer model we insert the weights $\lambda$ and $\lambda^{-1}$, for some column of bonds 
$\{((m_0,n);(m_0+1,n))\}$,
 alternately (Figure \ref{fig3}). We call such $m_0$-column a zipper, in accord with \cite{kenyondouble}. The shape of the zipper is in principle not important, this choice is about computability. For our purpose, $\lambda=\mathrm{e}^{i\beta}$ is suitable. Of course it will not necessarily lead to a probability measure for the dimer model
\footnote{It is nevertheless a probability measure in the case of a cylindrical domain with even vertical length ($N$ even) because of the balance between the number of dimers with $\lambda$ and $\lambda^{-1}$ fugacities in every dimer cover.}
, but it will be one for the double-dimer model if one dimer model has fugacities complex conjugate of those of the other, and if $\beta$ is sufficiently small. We do so, and check if the above requirements are met. Each nontrivial loop crosses the $m_0$-column odd number of times, and each trivial loop even number of times. It is clear that the contribution of doubled edges is then the factor $1$. So it is sufficient to show that the contribution of any loop crosses the $m_0$-column $l$ times, in each term in the expansion of the partition function is $\lambda^{\pm l(\text{mod}\hspace{.05cm}2)}$. It indeed holds because of the bipartiteness of the lattice, and the construction of the loops; each loop is a sequence of dimers, alternately belong to one of the contributing dimer models. If, for example, we move on a loop from one cross of the $m_0$-column made by the loop to the other, we will pass the set of weight $\{\lambda,\lambda^{-1}\}$ precisely. So the contribution of each loop only depends on the parity of the number of crosses of the $m_0$-column made by the loop. There is always two equiprobable possibilities of such contributions for each loop. The partition function is then the expectation of $(\frac{\mathrm{e}^{i\beta}+\mathrm{e}^{-i\beta}}{2})^k$ or equally $(\cos\beta)^k$, where $k$ indicates the number of loops around the annulus in each configuration of the double-dimer model on the cylinder. The reason for dividing by $2$ is that the weight of any double-dimer configuration with $l$ number of loops is proportional to $2^l$. If we orient the loops independently with equal probability, this expectation is somehow the discrete analog of the expectation of the layering operator, $e^{i\beta N_l}$, introduced in \cite{camia} for the Brownian Loop Soup (BLS) though the outcomes are very different in nature. 

In conclusion, we have the following identity,
\begin{align}
\langle(\cos\beta)^k\rangle_{dd}&=\mathcal{Z}_\beta/\mathcal{Z}_0\nonumber\\
&=\mathcal{Q}_\beta\mathcal{Q}_{-\beta}/\mathcal{Q}_0^2
\label{loopcylinder}
\end{align}
\begin{figure}[h]
	\includegraphics[width=.8\textwidth]{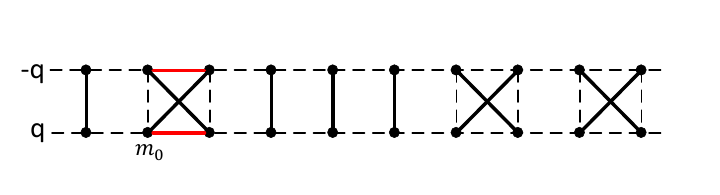}
	\centering
	\caption{An instance of coupled Fourier modes $q$ and
		$-q$ for the situation in Figure \ref{fig3}. The pair of red horizontal "dimers" is possible only in the $m_0$-column.}
	\label{fig4}
\end{figure}
where the index $dd$ indicates that the expectation is with respect to the probability measure of the double-dimer model, $\mathcal{Z}$ stands for the partition function of the resulting double-dimer model, $\mathcal{Q}_\beta$ is the partition function of the dimer model with the above weights, and $\mathcal{Q}_0$ is the partition function \eqref{cylinderpartitioneven} or \eqref{cylinderpartitionodd} depending on $N$ being even or odd (In the case $M$ odd and $N$ even, there is simply no nontrivial loop).

To compute the RHS of \eqref{loopcylinder}, we write the action for one contributing dimer model
\begin{align}\label{lambdaaction}
-\mathcal{S}_\beta = &\sum_n \{\sum_{m\neq m_0} c_{m+1,n}c_{m,n}+\sum_{m} (-1)^{m+1}c_{m,n+1}c_{m,n}\}\nonumber\\&+\sum_{n\hspace{.06cm}\text{odd}}\lambda c_{m_0+1,n}c_{m_0,n} + \sum_{n\hspace{.06cm}\text{even}}\lambda^{-1} c_{m_0+1,n}c_{m_0,n},\hspace{.8cm}\lambda=\mathrm{e}^{i\beta}
\end{align}
If the last two terms in \eqref{lambdaaction}
is written
$\sum_n (\cos\beta) c_{m_0+1,n}c_{m_0,n} + \sum_n (-1)^{n+1}(i\sin\beta) c_{m_0+1,n}c_{m_0,n}$, after Fourier transformation \eqref{partfourier} the action becomes
\begin{align}
-\mathcal{S}_\beta=\sum_{q=1}^N&\left\{\left[\sum_{m\neq m_0}-c_{m+1,q}c_{m,-q}-\sum_{m}i(-1)^{m+1}\cos\frac{\pi q}{N+1}c_{m,q}c_{m,-q}\right]\right.\nonumber\\
&\left.\vphantom{\left[\sum_{m\neq m_0}-c_{m+1,q}c_{m,-q}-\sum_{m}i(-1)^{m+1}\cos\frac{\pi q}{N+1}c_{m,q}c_{m,-q}\right]}\hspace{3.3cm}-\cos\beta\hspace{.1cm} c_{m_0+1,q}c_{m_0,-q} -i\sin\beta\hspace{.1cm} c_{m_0+1,q}c_{m_0,q}\right\}
\label{cylinderaction}
\end{align}

Here the argument is as before with a little change in the fugacities and form of the strips after Fourier transformation. On any cylindrical strip $[q,-q]$ there still exist cross-forms and vertical dimers, but just on the square $\{(m_0,q), (m_0+1,q), (m_0+1,-q), (m_0,-q)\}$ (let's call it $m_0$-square) there is an additional possibility of two horizontal dimers, and they always come together to make an admissible configuration. For either a cross-form or two horizontal dimers on $m_0$-square, the overall final factor
$1(=\cos^2\beta+\sin^2\beta)$ appears, the same as what occurs for a cross-form in the uniform case (Figure \ref{fig4}).
So the solution is also the same, except that now the two situations when a chain of tilted dimers occur, add the weight of $2\cos\beta$ (instead of $2$) to the solution for each strip. For $N$ even, the result is
\begin{align}
\mathcal{Q}_\beta=\prod_{q=1}^\frac{N}{2}\left[\left\{\sum_{p=0}^\frac{M}{2}\left[\left( \begin{array}{c} M-p \\ p \end{array} \right)+\left( \begin{array}{c} M-p-1 \\ p-1 \end{array} \right)\right](2\cos\frac{\pi q}{N+1})^{M-2p}\right\} + 2\cos\beta\right],\hspace{.3cm}(N\hspace{.1cm}\text{even})
\label{lambdacylinder}
\end{align}

For $N$ odd, $\mathcal{Q}_\beta$ does not yield a probability measure because of the single mode $q=\frac{N+1}{2}$ which produces the factor $(1+\mathrm{e}^{i\beta})$ in the partition function of the dimer model, 
\begin{align}
\mathcal{Q}_\beta=(1+\mathrm{e}^{i\beta})\prod_{q=1}^\frac{N-1}{2}\left[\left\{\sum_{p=0}^\frac{M}{2}\left[\left( \begin{array}{c} M-p \\ p \end{array} \right)+\left( \begin{array}{c} M-p-1 \\ p-1 \end{array} \right)\right](2\cos\frac{\pi q}{N+1})^{M-2p}\right\} + 2\cos\beta\right],\hspace{.3cm}(N\hspace{.1cm}\text{odd})
\end{align}
The resulting double-dimer partition function $\mathcal{Z}_\beta=\mathcal{Q}_\beta\mathcal{Q}_{-\beta}$ is nevertheless real. 

We should emphasize that here "a column of nontrivial weights" is more or less a matter of convenience and paves the way for the following sections. We could alternatively choose for $\lambda^{1/M}$ and $\lambda^{-1/M}$ fugacities, on odd and even rows of the cylinder respectively, and would eventually come to the same results; this is identical to the approach in \cite{kenyondouble} (up to the choice of Kasteleyn orientation) and could have the advantage of clarifying the relation to the associated Kasteleyn's matrix. 

For a very large lattice, things go on as in \cite{kenyondouble}, but we believe the result is
\begin{align}
\sum_{k=0}^{\infty}\mathbb{P}(k\hspace{.1cm}\text{loops})(\cos\beta)^k=\prod_{\substack{j=1\\ j\hspace{.1cm}\text{odd}}}^\infty\frac{(1+2\eta^j\cos\beta+\eta^{2j})^2}{(1+\eta^j)^4}
\label{discreteeven}
\end{align}
for $N$ even, and
\begin{align}
\sum_{k=0}^{\infty}\mathbb{P}(k\hspace{.1cm}\text{loops})(\cos\beta)^k=\frac{1+\cos\beta}{2}\prod_{\substack{j=2\\ j\hspace{.1cm}\text{even}}}^\infty\frac{(1+2\eta^j\cos\beta+\eta^{2j})^2}{(1+\eta^j)^4}
\label{discreteodd}
\end{align}
for $N$ odd, where $\eta=e^{-\frac{M}{2N}\pi}$. Here the LHS of \eqref{discreteeven} and \eqref{discreteodd} is the probability generating function of the number of nontrivial loops and is just $\langle(\cos\beta)^k\rangle_{dd}$, where "$\cos\beta$" serves as half of the "$X$" in \cite{kenyondouble}. On the other hand, if further $M\ll N$, we can somehow approximate the expectation \eqref{loopcylinder}
\begin{align}
\ln(\langle(\cos\beta)^k\rangle_{dd})\approx\frac{2N}{\pi}\int\ln(1-\frac{2\hspace{.1cm}b\hspace{.1cm}f(x)}{(1+f(x))^2})\frac{1}{\sqrt{1-x^2}}\hspace{.1cm}\dd x
\label{continuumannulus}
\end{align}
where $b=1-\cos\beta$, $x=\cos\frac{\pi q}{N+1}$, and $f(x)=(\sqrt{x^2+1}-x)^M$. Because of the intense concentration of the integrand near $x=0$, we can neglect $\frac{1}{\sqrt{1-x^2}}$ and replace $f(x)$ by $\frac{-f'(x)}{M}$. With differentiating \eqref{continuumannulus} with respect to $b$ (for the moment in the range where there is no singularity) and changing the order of integration, we obtain
\begin{align}
\sum_{k=0}^{\infty}\mathbb{P}(k\hspace{.1cm}\text{loops})(\cos\beta)^k\approx\exp\left({-\frac{N}{M}\frac{\beta^2}{\pi}}\right)
\label{cylinderlong}
\end{align}

The approximation \eqref{cylinderlong} can be obtained by a shortcut way, but as it happens, at the cost of restricting ourselves to low $\beta$. One can take the logarithm of the RHS of \eqref{discreteeven} to have the following approximation
\begin{align}
\ln(\langle(\cos\beta)^k\rangle_{dd})\approx\frac{2N}{M\pi}\int_0^\infty\ln\frac{1+2\cos\beta\hspace{.1cm}e^{-x}+e^{-2x}}{1+2\hspace{.1cm}e^{-x}+e^{-2x}}\hspace{.1cm}\dd x
\label{shortcontinuumannulus}
\end{align}
where $x$ is the continuum counterpart of $\eta^j$, which are distanced by $\frac{M\pi}{N}$. Now we assume that $\beta$ is small so we can expand the RHS of \eqref{shortcontinuumannulus} uniformly in $x$ to get the approximation \eqref{cylinderlong} from the following equality
\begin{align*}
\frac{2N}{M\pi}\int_0^\infty\ln\frac{-2\hspace{.1cm}e^{-x}\beta^2/2}{(1+e^{-x})^2}\hspace{.1cm}\dd x=-\frac{N}{M}\frac{\beta^2}{\pi}
\end{align*}

\begin{figure}[h]
	\includegraphics[width=.6\textwidth]{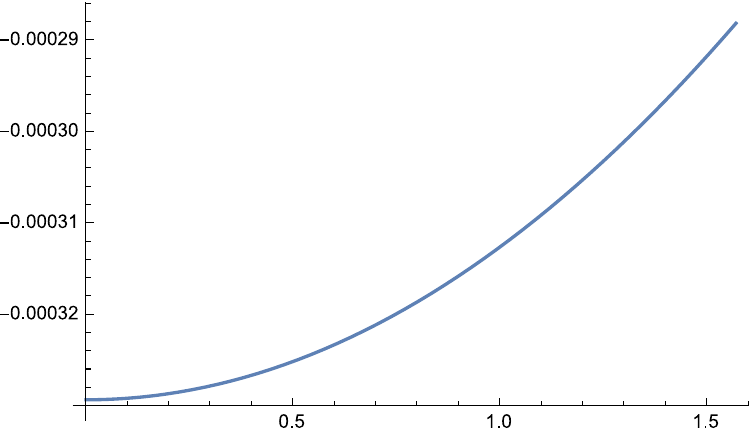}
	\centering
	\caption{The diagram of the relative error in the exponent of the RHS of \eqref{cylinderlong} as a function of $\beta$.}
	\label{relative}
\end{figure}

Figure \ref{relative} shows the diagram of the relative error 
$\frac{\text{RHS of \eqref{shortcontinuumannulus}}-\ln(\text{RHS of \eqref{lambdacylinder}})}{\ln(\text{RHS of \eqref{lambdacylinder}})}$ as a function of $\beta$, where $M=1000$, $N=10000$, and the product \eqref{lambdacylinder} is truncated at $j=1000$. However, one may note that the expression "$M\ll N$" is a bit formal and $\frac{M}{N}$, even up to $\frac{1}{2}$, is quite safe to get a very good approximation by \eqref{cylinderlong}.
 
To make an explicit connection with Coulomb gas formalism as well as check \eqref{cylinderlong}, we calculate a related observable which is very similar in quantity. The setup is as before, except that in each dimer model (but now $M$ and $N$ are odd), we insert a single monomer on the corner (see section \ref{monomer}), at the top-left side in one of them, and at the down-left side in the other, and further do likewise. There is always a path of dimers in the arising double-dimer model, which connects the top boundary of the cylinder to the bottom. In this case, the partition function of the double-dimer model leads to the distribution of the winding number of the path around the cylinder, that is
\begin{align}
\sum_{k=0}^{\infty}\mathbb{P}(k\hspace{.1cm}\text{windings})(\cos\beta)^k&=\mathcal{Z}_\beta\nonumber\\
&=\prod_{\substack{q=1\\ q\neq\frac{N+1}{2}}}^N\left[\left\{\sum_{p=0}^{\lfloor\frac{M}{2}\rfloor}\left[\left( \begin{array}{c} M-p \\ p \end{array} \right)+\left( \begin{array}{c} M-p-1 \\ p-1 \end{array} \right)\right](2\cos\frac{\pi q}{N+1})^{M-2p}\right\}+ 2\cos\beta\right]
\label{lambdacylinderwind}
\end{align}

Motivated by \cite{camia}, we call the LHS of \eqref{lambdacylinderwind} the expectation of the winding operator, $e^{i\beta N_w}$, where the orientation of each winding is considered with equal probability. For a very large lattice, we have
\begin{align}
\sum_{k=0}^{\infty}\mathbb{P}(k\hspace{.1cm}\text{windings})(\cos\beta)^k=\prod_{\substack{j=2\\ j\hspace{.1cm}\text{even}}}^\infty\frac{(1+2\eta^j\cos\beta+\eta^{2j})^2}{(1+\eta^j)^4}
\label{windingcylinder}
\end{align}
which for $M\ll N$ becomes
\begin{align}
\sum_{k=0}^{\infty}\mathbb{P}(k\hspace{.1cm}\text{windings})(\cos\beta)^k\approx\frac{2}{1+\cos\beta}\exp\left({-\frac{N}{M}\frac{\beta^2}{\pi}}\right)
\label{cylinderlongwind}
\end{align}
If we interpret the winding number in terms of height fluctuations in the dual picture, i.e. the corresponding height function defined on the dual lattice \cite{kenyondominoconf,kenyondomino}, the LHS of \eqref{cylinderlongwind} is the correlator of the exponential of the height, that is 
$\langle e^{iqh(r_1)}e^{-iqh(r_2)} \rangle$,
where the heights are taken to be integer multiples of $\pi$. The RHS of \eqref{cylinderlongwind} can be translated to $\exp(-2\pi x_q L/l)$, $x_q=q^2/{2g}$, which is the result obtained from the Coulomb gas method for the simple Gaussian model ($g=1$) on a very long cylinder ($L\gg l$) \cite{cardy}.
%=====================================================================%
\section{Loops surrounding two points}
\label{loopssurrounding}
We can use the the previous method to formally write the distribution of the number of loops surrounding an arbitrary number of faces (vertices of the dual) of the square lattice
$M\times N$. For two faces for example, assuming their centers to be located at $(M_1+\frac{1}{2},N_1+\frac{1}{2})$ and $(M_2+\frac{1}{2},N_2+\frac{1}{2})$, for one dimer model we insert the weights 
$\lambda_1$ and $\lambda_1^{-1}$, alternately on the column of bonds
$\{((M_1,n);(M_1+1,n))\}, n\leq N_1$, and likewise for another column replacing the index "1" by "2". We consider the conjugate of these weights for the other dimer model, and further assume that $M_1$ and $M_2$, as well as $N_1$ and $N_2$ have the same parity. 
\begin{figure}[h]
	\includegraphics[width=.8\textwidth]{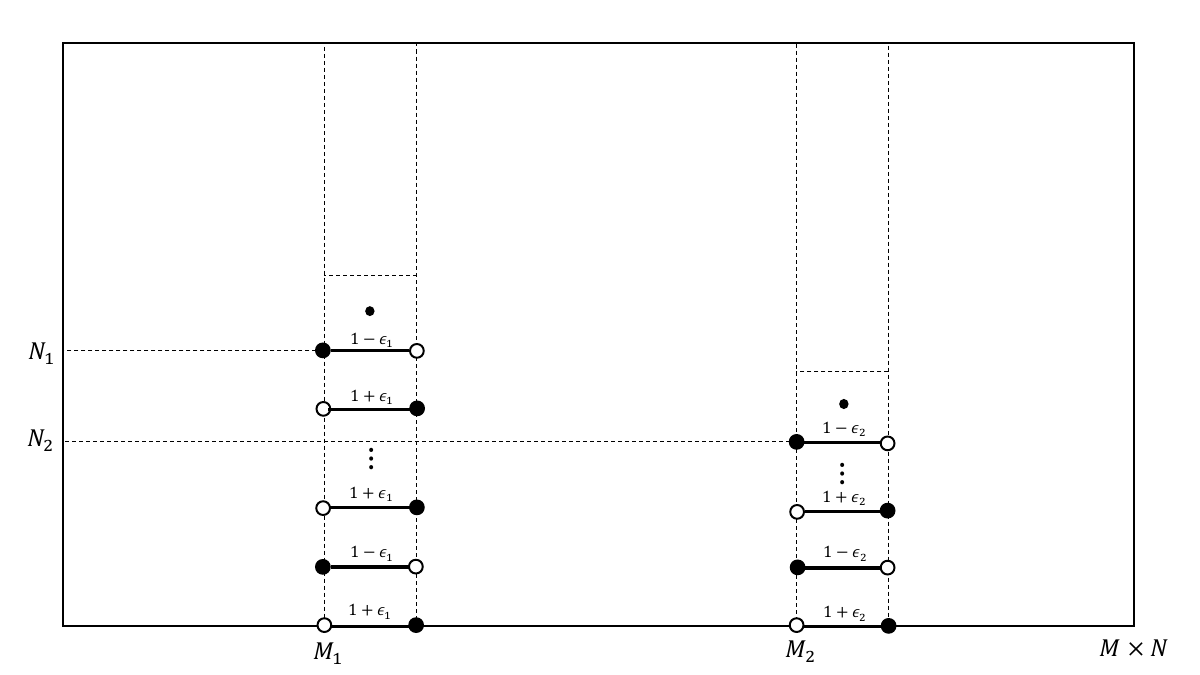}
	\centering
	\caption{The arrangement of fugacities in $M_1$- and $M_2$-columns, in one of the dimer models.}
	\label{fig5}
\end{figure}
Based on the arguments in the previous sections, one can show that in the expansion of the partition function of the double-dimer model, the coefficient of 
$(\cos(\beta_1+\beta_2))^k(\cos\beta_1)^{k_1}(\cos\beta_2)^{k_2}$ is proportional to the probability of occurring 
$k$ loops surrounding both points "1" and "2", $k_1$ loops around the point "1" but not "2", and $k_2$ loops around the point "2" but not "1".
In this case however, the complexity of the situation grows too fast with the exponents of $\lambda_1$ and $\lambda_2$ that a combinatorial analysis of the whole problem fails. In fact, even the previous decoupling to strips can not be met because of the lack of enough homogeneity in the model. Still, we can manage to compute the expectation of the number of loops surrounding the points. Let's slightly modify the weights above and replace
$\lambda_1$ (respectively $\lambda_1^{-1}$) with $1+\epsilon_1$ (respectively $1-\epsilon_1$), and do likewise for the index "2" (Figure \ref{fig5}).
If $\mathcal{Q}_{\epsilon_1,\epsilon_2}$ is the partition function of the associated dimer model and $\mathcal{Z}_{\epsilon_1,\epsilon_2}$ the corresponding one for the double-dimer model coming out of the identity
$\mathcal{Z}_{\epsilon_1,\epsilon_2}=\mathcal{Q}_{\epsilon_1,\epsilon_2}\mathcal{Q}_{-\epsilon_1,-\epsilon_2}$, the coefficient of 
$\epsilon_1\epsilon_2$ in the expansion of $\mathcal{Z}_{\epsilon_1,\epsilon_2}$ is then proportional to $\langle N_{12}\rangle$, the expectation of the number of loops surrounding both points "1" and "2", precisely
\begin{align}
\langle N_{12}\rangle=\frac{2C_{12}}{\mathcal{Q}_0}-\frac{2C_1C_2}{\mathcal{Q}_0^2}
\label{expect2point}
\end{align}
where $C_1$, $C_2$ and $C_{12}$ are determined by the following expansion 
\begin{align}
\mathcal{Q}_{\epsilon_1,\epsilon_2}=\mathcal{Q}_0+C_1\epsilon_1+C_2\epsilon_2+C_{12}\epsilon_1\epsilon_2+O(\epsilon_1^2)+O(\epsilon_2^2)
\label{partitionexpansion}
\end{align}

To compute these coefficients, we again do Fourier transformation \eqref{partfourier} for the Grassmannian action of the partition function $\mathcal{Q}_{\epsilon_1,\epsilon_2}$ with the free boundary conditions for Grassmann variables, $c_{M+1,n}=c_{m,N+1}=0$. The action becomes
\begin{align}
\label{actiontwopoint}
-\mathcal{S}_{\epsilon_1,\epsilon_2}=\sum_{m=1}^{M}\sum_{q=1}^N\{&-c_{m+1,q}c_{m,-q}-i(-1)^{m+1}\cos \frac{\pi q}{N+1}c_{m,q}c_{m,-q}\}\\\nonumber 
&-\epsilon_1\sum_{n_1=1}^{N_1}\frac{2}{N+1}\sum_{q,q'=1}^{N} c_{M_1+1,q}c_{M_1,q'}\sin\frac{\pi qn_1}{N+1}\sin\frac{\pi q'n_1}{N+1}\\\nonumber
&-\epsilon_2\sum_{n_2=1}^{N_1}\frac{2}{N+1}\sum_{q,q'=1}^{N} c_{M_2+1,q}c_{M_2,q'}\sin\frac{\pi qn_2}{N+1}\sin\frac{\pi q'n_2}{N+1}
\end{align}
The first row in \eqref{actiontwopoint} gives $\mathcal{Q}_0$ in terms of the multiplication of the contributions of coupled Fourier modes 
$q$ and $-q\equiv N+1-q$, that is \eqref{freepartitione}. We define $\mathcal{Q'}_{q_0}$ and $\mathcal{Q''}_{q_1,q_2}$ as $\mathcal{Q}_0$ in \eqref{freepartitione}, where contributions of the mode $q_0$, and modes $q_1$ and $q_2$ are excluded, respectively. 
For the coefficient $C_i$, one of the terms in the second or the third row in \eqref{actiontwopoint} contributes in configurations, that is, the weight of one of the bonds in the cross-form sitting in the $M_i$-square is changed to $\epsilon_i$. However, the style of couplings between Fourier modes will not change and the structure remains the same as before. The result is
\begin{align}
C_i=\sum_{n_i=1}^{N_i}\sum_{q_0=1}^\frac{N}{2}A_{q_0}^{(n_i)}\mathcal{Q'}_{q_0},\hspace{.5cm}i=1,2
\label{epsilon}
\end{align}
where by $A_{q_0}^{(n_i)}$ we mean
\begin{align*}
A_{q_0}^{(n_i)}=\frac{4}{N+1}(-1)^{n_i+1}\sin^2\frac{\pi q_0n_i}{N+1}\Theta_{i,q_0}
\end{align*}
\vspace{.1cm}
where $\Theta_{i,q_0}=\sum_{p=0}^\frac{M-2}{2}\left[\sum_{k=0}^p\left(\begin{array}{c} M_i-1-k \\ k \end{array}\right)\left(\begin{array}{c} M-M_i-1-(p-k) \\ p-k \end{array}\right)\right](2\cos\frac{\pi q_0}{N+1})^{M-2-2p}$.

\begin{figure}[h]
	\includegraphics[width=.8\textwidth]{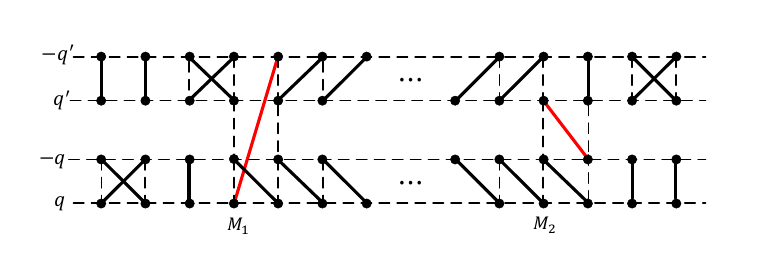}
	\centering
	\caption{An example of an "unusual" state, whose sign depends on both horizontal and vertical indices.}
	\label{fig6}
\end{figure}

\begin{figure}[h]
	\includegraphics[width=.8\textwidth]{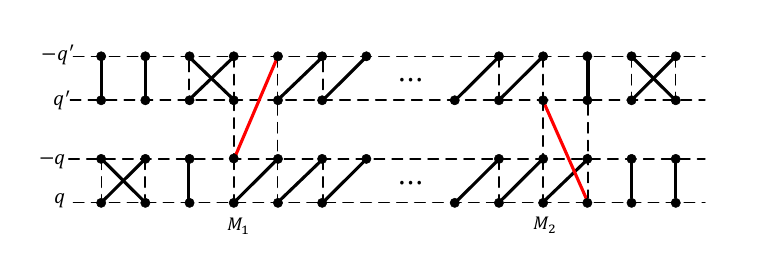}
	\centering
	\caption{An example of an "unusual" state, whose sign is independent of any index.}
	\label{fig7}
\end{figure}

\vspace{.1cm}
But the situation for the coefficient of $\epsilon_1\epsilon_2$ is more complicated; there can be terms supporting two "unusual" bonds between two different strips $[q_1,-q_1]$ and $[q_2,-q_2]$, and we have to be careful about the signs. Regarding this, the sign of a term, due to the contribution of two rows of the columns $\{((M_1,n);(M_1+1,n))\}$ and $\{((M_2,n);(M_2+1,n))\}$, naturally depends only on the parity of vertical indices of the rows with respect to each other. For an "unusual" state, either it also depends on the parity of horizontal indices of the columns with respect to each other (Figure \ref{fig6}), or it is independent of any index (Figure \ref{fig7}). The result can be expanded in three distinct kinds of terms
\begin{align}
C_{12}=\sum_{n_1=1}^{N_1}\sum_{n_2=1}^{N_2}\left\{\sum_{q_0=1}^\frac{N}{2}A^{(n_1,n_2)}_{q_0}\mathcal{Q'}_{q_0}
+\sum_{q_1\neq q_2}(A_{q_1,q_2}^{(n_1,n_2)}+B_{q_1,q_2}^{(n_1,n_2)})\mathcal{Q''}_{q_1,q_2}\right\}
\label{epsilon1epsilon2}
\end{align}
Along with some extra symbols, $A^{(n_1,n_2)}_{q_0}$, $A_{q_1,q_2}^{(n_1,n_2)}$ and $B_{q_1,q_2}^{(n_1,n_2)}$ are
\begin{align*}
A_{q_0}^{(n_1,n_2)}=(\frac{4}{N+1})^2(-1)^{n_1+n_2}\sin^2\frac{\pi q_0n_1}{N+1}\sin^2\frac{\pi q_0n_2}{N+1}\hspace{.1cm}\Psi_{q_0},
\end{align*}
\begin{align*}
A_{q_1,q_2}^{(n_1,n_2)}=(\frac{4}{N+1})^2(-1)^{n_1+n_2}\sin^2\frac{\pi q_1n_1}{N+1}\sin^2\frac{\pi q_2n_2}{N+1}\hspace{.1cm}\Xi_{q_1,q_2},
\end{align*}
\begin{align*}
B_{q_1,q_2}^{(n_1,n_2)}=(\frac{2}{N+1})^2\sin\frac{\pi q_1n_1}{N+1}\sin\frac{\pi q_2n_1}{N+1}\sin\frac{\pi q_1n_2}{N+1}&\sin\frac{\pi q_2n_2}{N+1}\hspace{.05cm}\left\{(-1)^{M_2-M_1}(-1)^{n_1+n_2}\sum_{\delta,\delta'}\Delta^{\delta,\delta'}_{q_1,q_2}\right.\\
&\left.\vphantom{(-1)^{M_2-M_1}(-1)^{n_1+n_2}\sum_{\delta,\delta'}\Delta^{\delta,\delta'}_{q_1,q_2}}\hspace{.9cm}+\sum_{\delta=\delta'}\Delta^{\delta,\delta'}_{q_1,q_2}-\sum_{\delta\neq\delta'}\Delta^{\delta,\delta'}_{q_1,q_2}\right\}.
\end{align*}
where $\delta$ and $\delta'$ can only take two values 0 and 1. Here $\Psi_{q_0}$, $\Xi_{q_1,q_2}$ and $\Delta^{\delta,\delta'}_{q_1,q_2}$ sum up the terms where both bonds with weights $\epsilon_1$ and $\epsilon_2$ are in the same strip, in two different strips, and connecting two different strips, respectively.  
\begin{align*}
\Psi_{q_0}=\sum_{p=0}^\frac{M-4}{2}&\left[\sum_{p_1+p_2+p_3=p}\left(\begin{array}{c} M_1-1-p_1 \\ p_1 \end{array}\right)\left(\begin{array}{c} M_2-M_1-2-p_2 \\ p_2 \end{array}\right)\left(\begin{array}{c} M-M_2-1-p_3 \\ p_3 \end{array}\right)\right]\\&\hspace{9.7cm}(2\cos\frac{\pi q_0}{N+1})^{M-4-2p},
\end{align*}
\begin{align*}
\Xi_{q_1,q_2}=&\left\{\sum_{p_1=0}^\frac{M-2}{2}\left[\sum_{k_1=0}^{p_1}\left(\begin{array}{c} M_1-1-k_1 \\ k_1 \end{array}\right)\left(\begin{array}{c} M-M_1-1-(p_1-k_1) \\ p_1-k_1 \end{array}\right)\right](2\cos\frac{\pi q_1}{N+1})^{M-2-2p_1}\right\}
\nonumber\\
&\left\{\sum_{p_2=0}^\frac{M-2}{2}\left[\sum_{k_2=0}^{p_2}\left(\begin{array}{c} M_2-1-k_2 \\ k_2 \end{array}\right)\left(\begin{array}{c} M-M_2-1-(p_2-k_2) \\ p_2-k_2 \end{array}\right)\right](2\cos\frac{\pi q_2}{N+1})^{M-2-2p_2}\right\},
\end{align*}
\begin{align}
\Delta_{q_1,q_2}^{\delta,\delta'}=&\left\{\sum_{p_1=0}^\frac{M-(M_2-M_1)-2+\delta+\delta'}{2}\left[\sum_{k_1=0}^{p_1}
\left(\begin{array}{c} M_1-1-k_1+\delta \\ k_1 \end{array}\right)\left(\begin{array}{c} M-M_2-1-(p_1-k_1)+\delta' \\ p_1-k_1 \end{array}\right)\right]\right.\nonumber\\&\left.\vphantom{\sum_{p_1=0}^\frac{M-(M_2-M_1)-2+\delta+\delta'}{2}\left[\sum_{k_1=0}^{p_1}
\left(\begin{array}{c} M_1-1-k_1+\delta \\ k_1 \end{array}\right)\left(\begin{array}{c} M-M_2-1-(p_1-k_1)+\delta' \\ p_1-k_1 \end{array}\right)\right]}\hspace{7.5cm}(2\cos\frac{\pi q_1}{N+1})^{M-(M_2-M_1)-2-2p_1+\delta+\delta'}\right\}\nonumber\\
&\left\{\sum_{p_2=0}^\frac{M-(M_2-M_1)-\delta-\delta'}{2}\left[\sum_{k_2=0}^{p_2}\left(\begin{array}{c} M_1-k_2-\delta \\ k_2 \end{array}\right)\left(\begin{array}{c} M-M_2-(p_2-k_2)-\delta' \\ p_2-k_2 \end{array}\right)\right]\right.\nonumber\\&\left.\vphantom{\sum_{p_2=0}^\frac{M-(M_2-M_1)-2+\delta'_1+\delta'_2}{2}\left[\sum_{k_2=0}^{p_2}\left(\begin{array}{c} M_1-1-k_2+\delta'_1 \\ k_2 \end{array}\right)\left(\begin{array}{c} M-M_2-1-(p_2-k_2)+\delta'_2 \\ p_2-k_2 \end{array}\right)\right]}\hspace{7.5cm}
(2\cos\frac{\pi q_2}{N+1})^{M-(M_2-M_1)-2p_2-\delta-\delta'}\right\}.
\label{ugly}
\end{align}
 Remembering the relation between the partition function of the dimer model and Chebyshev polynomials \eqref{chebyshevpartitionf}, the final result for $\langle N_{12}\rangle$ is briefly
 \begin{figure}[h]
 	\includegraphics[width=.8\textwidth]{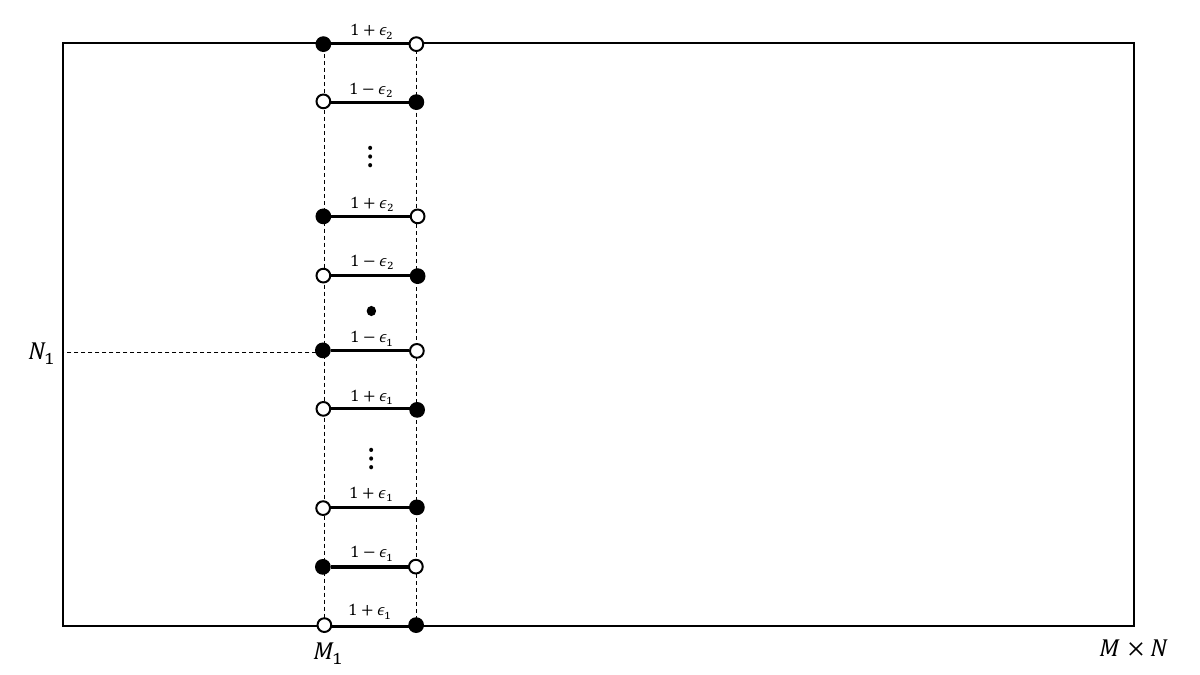}
 	\centering
 	\caption{The arrangement of fugacities in $M_1$-column, in one of the dimer models.}
 	\label{fig8}
 \end{figure}
\begin{align}
\langle N_{12}\rangle=\sum_{n_1=1}^{N_1}\sum_{n_2=1}^{N_2}\left\{2\sum_{q_0=1}^\frac{N}{2}(\frac{A^{(n_1,n_2)}_{q_0}}{U_M(q_0)}-\frac{A^{(n_1)}_{q_0}A^{(n_2)}_{q_0}}{U_M(q_0)^2})
+2\sum_{q_1\neq q_2}\frac{B_{q_1,q_2}^{(n_1,n_2)}}{U_M(q_1)U_M(q_2)}\right\}
\label{N12}
\end{align}
along with a little abuse of notation $U_M(q)\equiv |U_M(i\cos\frac{\pi q}{N+1})|$, where $U_M(x)$ is the M-th Chebyshev polynomial of the second kind. 

Another interesting situation, motivated by \cite{camia}, is the so-called charge conservation condition, where we are interested in $\langle N_{12'}+N_{1'2}\rangle$, the expectation of the number of loops surrounding point "1" but not "2" and point "2" but not "1". To make the situation clear, we first obtain $\langle N^\odot_i\rangle$, the expectation of the number of loops around one face. If we consider the above arrangement of fugacities (in both dimer models) only for one of the faces, say $(M_1+\frac{1}{2},N_1+\frac{1}{2})$, a zero coefficient in the first order of $\epsilon_1$ emerges in the expansion of the partition function of the double-dimer model. To avoid this, we insert fugacities $1\mp\epsilon_2$ on the bonds remained in the same column in one dimer model (Figure \ref{fig8}), and likewise in the other replacing $\pm$ by $\mp$ and vice versa. For our purpose, the same is obtained if instead of two upward and downward $M_1$-columns terminating in $(M_1+\frac{1}{2},N_1+\frac{1}{2})$-face, we merely replace $1\pm\epsilon_1$ in one dimer model (respectively $1\mp\epsilon_1$ in the other) by $(1\pm\epsilon_1)(1\pm\epsilon_2)$ (respectively $(1\mp\epsilon_1)(1\mp\epsilon_2)$) and look at the coefficient of $\epsilon_1\epsilon_2$ in the expansion. The desired expectation is
\begin{align}
\langle N^\odot_1\rangle=2(\frac{C_1}{\mathcal{Q}_0})^2
\label{expect1point}
\end{align}
in terms of the above notations.

\begin{figure}[h]
	\includegraphics[width=.8\textwidth]{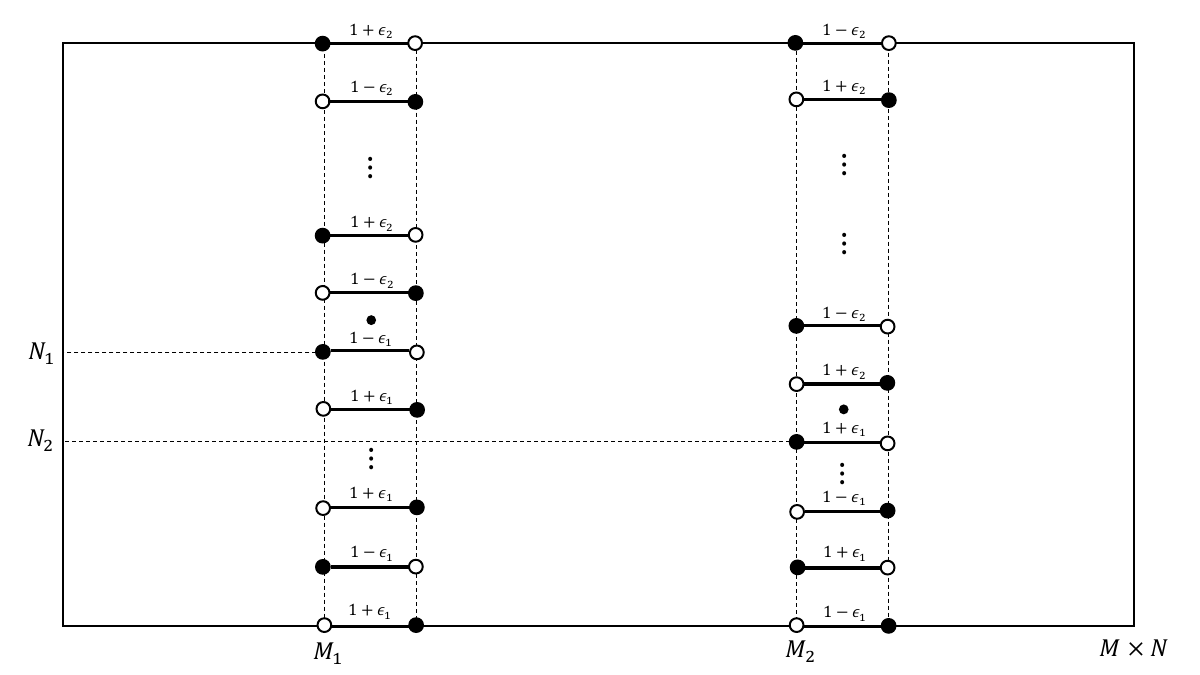}
	\centering
	\caption{The arrangement of fugacities in $M_1$- and $M_2$-column, in one of the dimer models.}
	\label{fig9}
\end{figure}

Suppose now we insert the aforementioned column of fugacities also for the other face, but replace $\epsilon_i$ by $-\epsilon_i$ (Figure \ref{fig9}). This arrangement of fugacities, again for our purposes, is identical to replacing $1\pm\epsilon_1$ by $(1\pm\epsilon_1)(1\pm\epsilon_2)$ in both columns (and a similar approach as before in the other dimer model). The coefficient of $\epsilon_1\epsilon_2$ is the expectation $\langle N_{12'}+N_{1'2}\rangle$, which is
\begin{align}
\langle N_{12'}+N_{1'2}\rangle=2(\frac{C_1+C_2}{\mathcal{Q}_0})^2-4\frac{C_{12}}{\mathcal{Q}_0}
\label{chargeconservation}
\end{align}
We can see that the results \eqref{expect2point}, \eqref{expect1point} and \eqref{chargeconservation} are simultaneously consistent with each other. 

With careful arranging of fugacities, many other observables such as generalizations of the above quantities to an arbitrary number of faces, the probability of some given dimers belonging to the same loop, n-point functions of dimer correlations in the dimer model, etc. can be computed similarly. 
For instance, the probability of two dimers, $\{(M_1,N_1);(M_1+1,N_1)\}$ and $\{(M_2,N_2);(M_2+1,N_2)\}$, belonging to the same loop can be obtained by inserting the fugacities $1+\epsilon_1$ and $1+\epsilon_2$ on the corresponding bonds respectively, in one dimer model, and likewise replacing $\epsilon_i$ by $-\epsilon_i$ in the other. The desired probability is obtained from \eqref{N12} by considering only the "$n_1=N_1,n_2=N_2$" term in the summation "$\sum_{n_1}\sum_{n_2}$". 
This result would be expected since the expectation $\langle N_{12}\rangle$ is the summation of all such probabilities for relevant dimers.

After all, we are interested in the scaling limits of such observables \footnote{Though there is delicacy in the choice of the boundary conditions \cite{kenyondouble} and we have to be careful in taking the continuum limit, here we will not concern ourselves with the issue.}. Here we try to approximate \eqref{N12} for very large lattices while keeping $\frac{M}{N}$, $\frac{M_1}{M}$, $\frac{M_2}{M}$, $\frac{N_1}{N}$ and $\frac{N_2}{N}$ fixed. This case is especially interesting. First, we rewrite the expressions in \eqref{ugly}, again using Chebyshev polynomials, for convenience and more insight into the result
\begin{align}
&\Theta_{i,q_0}=U_{M_i-1}(q_0)U_{M-M_i-1}(q_0),\nonumber\\
&\Psi_{q_0}=U_{M_1-1}(q_0)U_{M_2-M_1-2}(q_0)U_{M-M_2-1}(q_0),\nonumber\\
&\Delta_{q_1,q_2}^{\delta,\delta'}=U_{M_1-1+\delta}(q_1)U_{M-M_2-1+\delta'}(q_1)U_{M_1-\delta}(q_2)U_{M-M_2-\delta'}(q_2)
\end{align}
Now we note that each term in \eqref{N12} is only significant when $\cos\frac{\pi q_i}{N+1}\approx 0$. On the other hand, for small $\epsilon>0$ and $\theta=\frac{\pi}{2}+\epsilon$, $\ln(x+\sqrt{1+x^2})=\sinh^{-1} x$ leads to $\sqrt{1+\cos^2\theta}+\cos\theta=\exp(\epsilon+O(\epsilon^3))$. Regarding this and the identity $U_n(q)=(2\sqrt{1+\cos^2\theta})^{-1}[(\sqrt{1+\cos^2\theta}+\cos\theta)^{n+1}-(-\sqrt{1+\cos^2\theta}+\cos\theta)^{n+1}]$, $\theta=\frac{\pi q}{N+1}$, for large $n$ we conclude that \footnote{Indeed this approximation has been used to obtain \eqref{discreteeven} and \eqref{discreteodd} in the previous section, as done in \cite{kenyondouble}.} 
\begin{align}
U_n(q)\approx\left\{\begin{array}{rcl} \sinh(n\epsilon) & n\hspace{.1cm}\mbox{odd}\\ \cosh(n\epsilon)  & n\hspace{.1cm}\mbox{even}\end{array}\right.
\label{approximation}
\end{align}
Without loss of generality, we assume that $M_1$ and $M_2>M_1$ are both even, so for example $U_M(q)\approx\cosh M\epsilon$, $U_{M_2-M_1-2}(q)\approx\cosh(M_2-M_1)\epsilon$, $U_{M-M_1-1}(q)\approx\sinh(M-M_1)\epsilon$ and so on. These approximations will be exact in the thermodynamic limit. There are also some coefficients in front of $\Theta_{q_0}$, $\Psi_{q_0}$ and $\Delta^{\delta,\delta'}_{q_1,q_2}$ in the expansion of \eqref{N12} which are translated into suitable expressions. Finally, with the aid of the identities \eqref{cos} and \eqref{-cos}, and using $\sum_{\epsilon}$ formally at the moment, we sum over $n_1$ and $n_2$ to obtain \footnote{Many necessary parentheses are ignored to make the appearance of the expressions more pleasant. We hope there will be no ambiguity about this.}
 
\begin{align}
&\langle N_{12}\rangle\approx\frac{1}{2N^2}\sum_{\epsilon_0}\left\{\frac{\sin(2N_1)\epsilon_0\sin(2N_2)\epsilon_0}{\sin^2\epsilon_0}\hspace{.1cm}\frac{\cosh(M-M_2+M_1)(2\epsilon_0)-\cosh(M-M_2-M_1)(2\epsilon_0)}{\cosh (2M\epsilon_0)+1}\right\}\nonumber\\
&+\frac{1}{2N^2}\sum_{\epsilon_1\neq\epsilon_2}\left\{\frac{\sin(2N_1)\frac{\epsilon_1+\epsilon_2}{2}\sin(2N_2)\frac{\epsilon_1+\epsilon_2}{2}}{\sin^2\frac{\epsilon_1+\epsilon_2}{2}}\right.\nonumber\\
&\left.\vphantom{\frac{\sin(2N_1+1)(\frac{\epsilon_1+\epsilon_2}{2})\sin(2N_2+1)(\frac{\epsilon_1+\epsilon_2}{2})}{\sin^2\frac{\epsilon_1+\epsilon_2}{2}}}\hspace{5cm}\frac{\cosh(M-M_2+M_1)(\epsilon_1+\epsilon_2)-\cosh(M-M_2-M_1)(\epsilon_1+\epsilon_2)}{\cosh M(\epsilon_1+\epsilon_2)+\cosh M(\epsilon_1-\epsilon_2)}\right.\nonumber\\
&\left.\vphantom{\frac{\sin(2N_1+1)\frac{\epsilon_1+\epsilon_2}{2}\sin(2N_2+1)\frac{\epsilon_1+\epsilon_2}{2}}{\sin^2\frac{\epsilon_1+\epsilon_2}{2}}}\hspace{.1cm}+\frac{\sin(2N_1)\frac{\epsilon_1-\epsilon_2}{2}\sin(2N_2)\frac{\epsilon_1-\epsilon_2}{2}}{\sin^2\frac{\epsilon_1-\epsilon_2}{2}}
\hspace{.1cm}\frac{\cosh(M-M_2+M_1)(\epsilon_1-\epsilon_2)-\cosh(M-M_2-M_1)(\epsilon_1-\epsilon_2)}{\cosh M(\epsilon_1+\epsilon_2)+\cosh M(\epsilon_1-\epsilon_2)}\right\}
\label{approx}
\end{align}
where we choose the index "$i$" of the associated $\epsilon$ according to the mode $q_i$. Many terms have disappeared in appreciation of larger and larger lattices. 

Now we come to the point of identifying $\epsilon_i$'s. Take $q=\frac{N}{2}+k$ then $\epsilon=\frac{\pi(2k-1)}{2(N+1)}$. So $\epsilon_1+\epsilon_2=\frac{\pi(k_1+k_2-1)}{(N+1)}$ and $\epsilon_1-\epsilon_2=\frac{\pi(k_1-k_2)}{(N+1)}$. If we change the role of $\epsilon_1+\epsilon_2$ and $\epsilon_1-\epsilon_2$  in the second term of the second summation, \eqref{approx} becomes
\begin{align}
\langle N_{12}\rangle\approx\sum_{j=1}^\infty&\left\{\frac{2}{\pi^2j^2}\sin(\frac{N_1}{N}\pi j)\sin(\frac{N_2}{N}\pi j)\left[\cosh(\frac{M-M_2+M_1}{N}\pi j)-\cosh(\frac{M-M_2-M_1}{N}\pi j)\right]\right.\nonumber\\
&\left.\vphantom{\frac{1}{\pi^2j^2}\sin(\frac{N_1}{N}\pi j)\sin(\frac{N_2}{N}\pi j)\left[\cosh\frac{M-M_2+M_1}{N}\pi j-\cosh(M-M_2-M_1)j\right]}\hspace{6.7cm}\sum_{k=-\infty}^{\infty}\frac{1}{\cosh(\frac{M}{N}\pi j)+\cosh(\frac{M}{N}\pi k)}\delta_{\frac{j+k-1}{2},0}\right\}
\label{exact}
\end{align}
where we use the approximation $\sin\epsilon\approx\epsilon$ for small $\epsilon$. From \eqref{asli} and \eqref{cosh}, we conclude that in the thermodynamic limit
\begin{align}
\langle N_{12}\rangle=\frac{8}{\pi^3}\frac{M}{N}\sum_{m,n=1}^\infty\frac{\sin(\frac{M_1}{M}m\pi)\sin(\frac{N_1}{N}n\pi)\sin(\frac{M_2}{M}m\pi)\sin(\frac{N_2}{N}n\pi)}{m^2+(\frac{M}{N})^2n^2}
\label{final}
\end{align}
This is directly proportional (with proportionality constant $\frac{2}{\pi}$) to the Green's function for the Dirichlet problem posed for the Laplace equation on any rectangle $\{0<x<a, 0<y<b; \frac{a}{b}=\frac{M}{N}\}$ \cite{pinsky}. The result \eqref{final} is absolutely expected if we assume the relation between the double-dimer loop ensemble and CLE$_4$ \cite{milwatwil}.
%==========================================================================================
%
\section{Revisiting boundary monomer-dimer problem}
\label{monomer}
We somewhat digress here and make use of the previous approach to simply solve the monomer-dimer problem where it is exactly solvable; the case of single monomer on the corner is relevant to the next section.

The inclusion of monomers is a priori equivalent to the insertion of magnetic fields at the locations of defect \cite{allegra}. While this interpretation provides us with a noticeable physical insight into the monomer-dimer situation, as dislocality does not emerge in the boundary case (or some particular situations, see below), there's no need to apply auxiliary variables representing magnetic fields, in our half-combinatorial approach; we directly compute the correlation of monomers, which is simply defined as the ratio between partition functions with and without defects. 

Assume for example that there are two monomers on the boundary, on the same line at positions 
$(m_1,1)$ and
$(m_2,1)$, $m_1 <m_2$.
If we use the Fourier transformation \eqref{partfourier}, each permissible configuration has two of the Fourier transformed variables sticking to some strip
$[q,-q]$. This means that (other than some factors) an extra multiplicative term of kind 
$c_{m_1,q}c_{m_2,-q}$, $q=1,...,N$, appears in front of every (valid) term in the Fourier transformed expansion of the partition function of close-packed dimer case.
\begin{figure}[h]
	\includegraphics[width=1\textwidth]{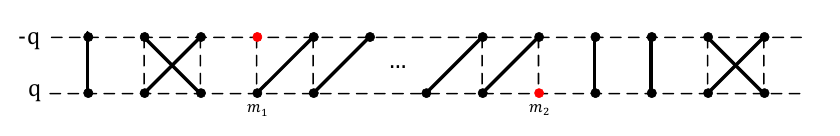}
	\centering
	\caption{An instance of coupled Fourier modes $q$ and $-q$ in the presence of two monomers (in red) on the boundary.}
	\label{fig10}
\end{figure} 
Figure \ref{fig10} shows an instance of such configuration in the momentum space, where the weight of such state has extra factor $(-1)^{m_1-m_2+1}$, with respect to the other situation where the monomers at $m_1$ and $m_2$ are respectively on "$q$" and "$-q$" lines. This implies that the partition function of the corresponding dimer model vanishes when $m_1$ and $m_2$ have the same parity. The solution is:
\begin{align}
\mathcal{Q}_{m_1,m_2}=\sum_{q_0=1}^\frac{N}{2}A_{q_0}^\parallel\mathcal{Q'}_{q_0}
\end{align}
where by 
$A^\parallel_{q_0}$
we mean
\begin{align}
A^\parallel_{q_0}=\frac{2(1-(-1)^{m_2-m_1})}{N+1}\sin^2\frac{\pi q_0}{N+1}\sum_{p=0}^\frac{M-(m_2-m_1)-1}{2}&\left[\sum_{k=0}^p\left(\begin{array}{c} m_1-1-k \\ k \end{array}\right)\left(\begin{array}{c} M-m_2-(p-k) \\ p-k \end{array}\right)\right]\nonumber\\&\hspace{2.5cm}(2\cos\frac{\pi q_0}{N+1})^{M-(m_2-m_1)-1-2p}
\end{align} 

For two monomers on the boundary, not inline and at positions $(m_0,1)$ and $(M,n_0)$, the partition function is:
\begin{align}
\mathcal{Q}_{m_0,n_0}=\sum_{q_0=1}^\frac{N}{2}A_{q_0}^\perp\mathcal{Q'}_{q_0}
\end{align}
where $A_{q_0}^\perp$ is:
\begin{align}
A_{q_0}^\perp=\frac{2(1-(-1)^{m_0})}{N+1}\sin\frac{\pi q_0}{N+1}\sin\frac{\pi q_0n_0}{N+1}\left[\sum_{p=0}^\frac{m_0-1}{2}\left(\begin{array}{c} m_0-1-p \\ p \end{array}\right)(2\cos\frac{\pi q_0}{N+1})^{m_0-1-2p}\right]
\end{align}

It is straightforward to compute the partition function of the dimer model in the presence of $2n$ number of monomers on the boundary. For example, if there are four inline monomers at $(m_i,1),\hspace{.1cm}i=1,...,4$ and $m_i<m_j$ for $i<j$, then:
\begin{align}
\mathcal{Q}_{m_1,m_2,m_3,m_4}=\sum_{q_0}A_{q_0}^{1,2,3,4}\hspace{.05cm}\mathcal{Q'}_{q_0}+\sum_{q_1\neq q_2}(A_{q_1,q_2}^{1,2;3,4}-A_{q_1,q_2}^{1,3;2,4}+A_{q_1,q_2}^{1,4;2,3})\hspace{.05cm}\mathcal{Q''}_{q_1,q_2}
\label{4monomer}
\end{align}
where $A_{q_0}^{1,2,3,4}$ accounts for the condition where all monomers stick to the same strip, and $A_{q_1,q_2}^{1,2;3,4}$, $A_{q_1,q_2}^{1,3;2,4}$ and $A_{q_1,q_2}^{1,4;2,3}$ sum up the contributions of the situations where two pairs of the monomers stick to two strips (which is reserved to which one is clarified by the indices). The amount of these quantities are as the following:
\begin{align*}
A_{q_0}^{1,2,3,4}=&(\frac{2}{N+1})^2(1-(-1)^{m_2-m_1})(1-(-1)^{m_4-m_3})\sin^4\frac{\pi q_0}{N+1}\nonumber\\&\sum_{p=0}^\frac{M-(m_4-m_3)-(m_2-m_1)-2}{2}\left[\sum_{p_1+p_2+p_3=p}\left(\begin{array}{c} m_1-1-p_1 \\ p_1 \end{array}\right)\left(\begin{array}{c} m_3-m_2-1-p_2 \\ p_2 \end{array}\right)
\left(\begin{array}{c} M-m_4-p_3 \\ p_3 \end{array}\right)\right]\nonumber\\&(2\cos\frac{\pi q_0}{N+1})^{M-(m_4-m_3)-(m_2-m_1)-2-2p}
\end{align*}
\begin{align*}
A_{q_1,q_2}^{1,2;3,4}=&(\frac{2}{N+1})^2(1-(-1)^{m_2-m_1})(1-(-1)^{m_4-m_3})\sin^2\frac{\pi q_1}{N+1}\sin^2\frac{\pi q_2}{N+1}\nonumber\\
&\left\{\sum_{p=0}^\frac{M-(m_2-m_1)-1}{2}\left[\sum_{k=0}^p\left(\begin{array}{c} m_1-1-k \\ k \end{array}\right)\left(\begin{array}{c} M-m_2-(p-k) \\ p-k \end{array}\right)\right](2\cos\frac{\pi q_0}{N+1})^{M-(m_2-m_1)-1-2p}\right\}\nonumber\\
&\left\{\sum_{p=0}^\frac{M-(m_4-m_3)-1}{2}\left[\sum_{k=0}^p\left(\begin{array}{c} m_3-1-k \\ k \end{array}\right)\left(\begin{array}{c} M-m_4-(p-k) \\ p-k \end{array}\right)\right](2\cos\frac{\pi q_0}{N+1})^{M-(m_4-m_3)-1-2p}\right\}
\end{align*},
\begin{align*}
A_{q_1,q_2}^{1,3;2,4}=&(\frac{2}{N+1})^2(1-(-1)^{m_3-m_1})(1-(-1)^{m_4-m_2})\sin^2\frac{\pi q_1}{N+1}\sin^2\frac{\pi q_2}{N+1}\nonumber\\
&\left\{\sum_{p=0}^\frac{M-(m_3-m_1)-1}{2}\left[\sum_{k=0}^p\left(\begin{array}{c} m_1-1-k \\ k \end{array}\right)\left(\begin{array}{c} M-m_3-(p-k) \\ p-k \end{array}\right)\right](2\cos\frac{\pi q_0}{N+1})^{M-(m_3-m_1)-1-2p}\right\}\nonumber\\
&\left\{\sum_{p=0}^\frac{M-(m_4-m_2)-1}{2}\left[\sum_{k=0}^p\left(\begin{array}{c} m_2-1-k \\ k \end{array}\right)\left(\begin{array}{c} M-m_4-(p-k) \\ p-k \end{array}\right)\right](2\cos\frac{\pi q_0}{N+1})^{M-(m_4-m_2)-1-2p}\right\}\nonumber
\end{align*}
and
\begin{align*}
A_{q_1,q_2}^{1,4;2,3}=&(\frac{2}{N+1})^2(1-(-1)^{m_4-m_1})(1-(-1)^{m_3-m_2})\sin^2\frac{\pi q_1}{N+1}\sin^2\frac{\pi q_2}{N+1}\nonumber\\
&\left\{\sum_{p=0}^\frac{M-(m_4-m_1)-1}{2}\left[\sum_{k=0}^p\left(\begin{array}{c} m_1-1-k \\ k \end{array}\right)\left(\begin{array}{c} M-m_4-(p-k) \\ p-k \end{array}\right)\right](2\cos\frac{\pi q_0}{N+1})^{M-(m_4-m_1)-1-2p}\right\}\nonumber\\
&\left\{\sum_{p=0}^\frac{M-(m_3-m_2)-1}{2}\left[\sum_{k=0}^p\left(\begin{array}{c} m_2-1-k \\ k \end{array}\right)\left(\begin{array}{c} M-m_3-(p-k) \\ p-k \end{array}\right)\right](2\cos\frac{\pi q_0}{N+1})^{M-(m_3-m_2)-1-2p}\right\}
\end{align*}

The minus sign in front of $A_{q_1,q_2}^{1,3;2,4}$ in \eqref{4monomer} is due to the transposition of $c_{{m_2},1}$ and $c_{{m_3},1}$ in the multiplicative term $c_{{m_1},1}c_{{m_2},1}c_{{m_3},1}c_{{m_4},1}$ in the expansion of the partition function of the dimer model in the presence of four inline monomers. 

For single monomer on the boundary, when $M$ and $N$ are both odd:
\begin{align*}
\mathcal{Q}_{1}=\sqrt{\frac{2}{N+1}}i^{n}\sin\frac{\pi\frac{N+1}{2}n}{N+1}i^\frac{N-1}{2}\prod_{q=1}^{\frac{N-1}{2}}\left[\sum_{p=0}^\frac{M-1}{2}\left(\begin{array}{c} M-p \\ p \end{array}\right)(2\cos\frac{\pi q}{N+1})^{M-2p}\right]
\end{align*}
which is zero for $n$ even and non-zero $n$-independent for $n$ odd, that is
\begin{align}
\mathcal{Q}_{1}=\sqrt{\frac{2}{N+1}}\prod_{q=1}^{\frac{N-1}{2}}\left[\sum_{p=0}^\frac{M-1}{2}\left(\begin{array}{c} M-p \\ p \end{array}\right)(2\cos\frac{\pi q}{N+1})^{M-2p}\right]
\label{singlemonomer}
\end{align}
regardless of the sign or $i$ factor (which is independent of $n$ anyway).
We can show that \eqref{singlemonomer} is identical to the corresponding result in 
\cite{allegraexact}
with the aid of the identities \eqref{prodlarge} and \eqref{prodsmall}.

There are situations where the monomer correlations can obviously be interpreted as dimer ones, i.e. the collection of monomers is identical to a union of some (perhaps bulk) dimers. In these cases, methods similar to those of the previous section are readily applicable. 
%==========================================================================================%
\section{Left-passage probability}
\label{leftpassage}
\begin{figure}[h]
	\includegraphics[width=.8\textwidth]{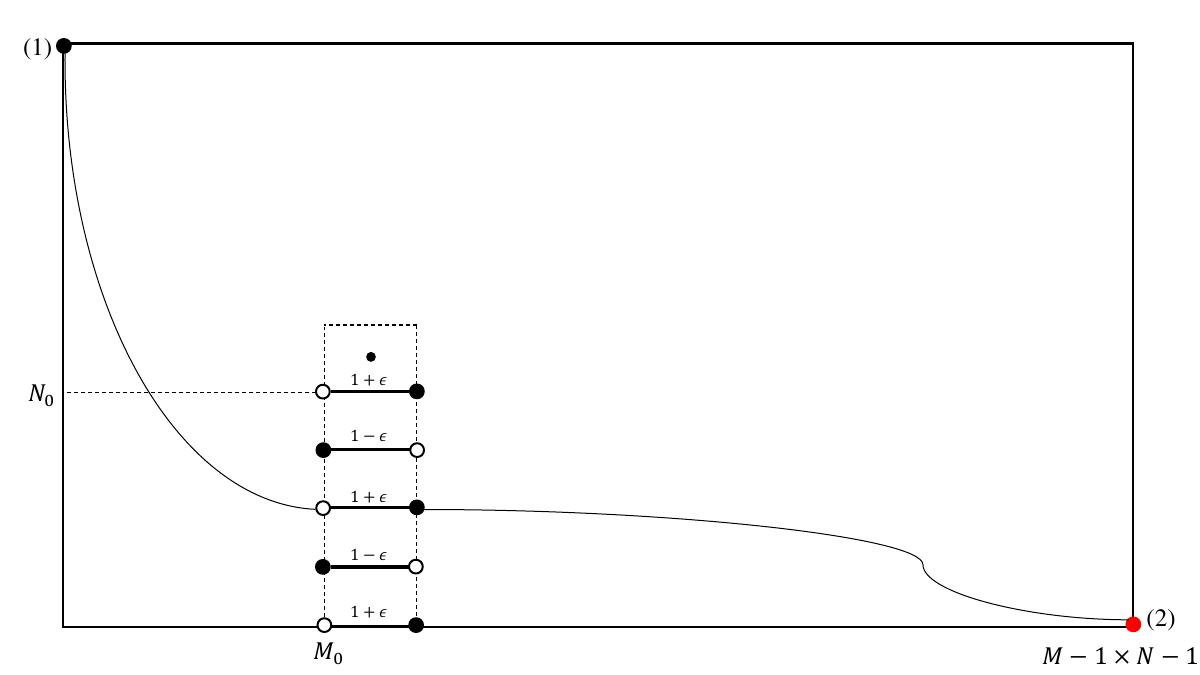}
	\centering
	\caption{A path due to the insertion of one monomer on the top-left corner in the first dimer model, and another (in red) on the bottom-right corner in the second one.}
	\label{fig11}
\end{figure}

Assume boundary conditions such that there exists a chordal path between two vertices on the boundary, in the configurations of the double-dimer model on a $M\times N$ square lattice. We can do that, for example, by fixing dimers on two adjacent sides of the lattice, alternately in both underlying dimer configurations. For $M$ and $N$ even, this is equivalent to "filling" these sides with monomers and then, inserting one monomer in either dimer configuration, one at the top-left and the other at the bottom-right corner of the remained $M-1\times N-1$ square lattice (Figure \ref{fig11}). With this "wired/free" boundary conditions, there always be a path from one corner to the other in the double-dimer configurations, and we want to compute the probability of the loop completing this path encompassing an arbitrary face of the lattice, i.e. the left-passage probability. Similar to the previous computation in Sec. \ref{loopssurrounding}, we insert the weights 
$1+\epsilon$ and $1-\epsilon$, alternately on the column of bonds
$\{((M_0,n);(M_0+1,n))\}, n\leq N_0$, for one dimer model, and likewise for the other replacing $1+\epsilon$ by $1-\epsilon$, and vice versa. The actions are
\begin{align}
\label{actionleftpassage}
-\mathcal{S}_{\epsilon}=\sum_{m=1}^{M-1}\sum_{q=1}^{N-1}\{&-c_{m+1,q}c_{m,-q}-i(-1)^{m+1}\cos \frac{\pi q}{N}c_{m,q}c_{m,-q}\}\nonumber\\
&-\epsilon\frac{2}{N}\sum_{n=1}^{N_0}\sum_{q,q'=1}^{N-1} c_{M_0+1,q}c_{M_0,q'}\sin\frac{\pi qn}{N}\sin\frac{\pi q'n}{N}
\end{align}
in the former case and $\mathcal{S}_{-\epsilon}$ in the latter.
We denote by $\mathcal{Z}_{\epsilon}$ the partition function of the double-dimer model coming out of the identity
$\mathcal{Z}_{\epsilon}=\mathcal{Q}^{(1)}_{\epsilon}\mathcal{Q}^{(2)}_{-\epsilon}$, where $\mathcal{Q}^{(1)}_{\epsilon}$ and $\mathcal{Q}^{(2)}_{-\epsilon}$ are partition functions of the contributing dimer models, equivalently integrations of the actions $\mathcal{S}_{\epsilon}$ and $\mathcal{S}_{-\epsilon}$ in the presence of one monomer, at the top-left and down-right corner, respectively.  
\begin{align}
\mathcal{Q}_{\epsilon}^{(1)}=\mathcal{Q}_0^{(1)}+\epsilon C^{(1)}+O(\epsilon^2),\hspace{1cm}
\mathcal{Q}_{\epsilon}^{(2)}=\mathcal{Q}_0^{(2)}+\epsilon C^{(2)}+O(\epsilon^2)
\label{partitionleftpassage}
\end{align}
and
\begin{align}
\mathcal{Q}_0^{(1)}&=\sqrt{\frac{2}{N}}\prod_{q=1}^{\frac{N-2}{2}}\left[\sum_{p=0}^{\left\lfloor{\frac{M-1}{2}}\right\rfloor}\left( \begin{array}{c} M-1-p \\ p \end{array} \right)(2\cos\frac{\pi q}{N})^{M-1-2p}\right]\nonumber\\
&=\mathcal{Q}_0^{(2)}
\end{align}

The coefficient of 
$\epsilon$ in the expansion of $\mathcal{Z}_{\epsilon}$ is then proportional to the aforementioned probability for the face $z=(M_0+\frac{1}{2},N_0+\frac{1}{2})$, that is
\begin{align*}
\mathbb{P}_\text{left-passage}(z)&=\text{the coefficient of $\epsilon$ in the expansion of $\frac{\mathcal{Q}_{\epsilon}^{(1)}\mathcal{Q}_{\epsilon}^{(2)}}{\mathcal{Q}_{0}^{(1)}\mathcal{Q}_{0}^{(2)}}$}\nonumber\\&=\frac{C^{(1)}+C^{(2)}}{\mathcal{Q}_{0}}
\end{align*}
where $\mathcal{Q}_0=\mathcal{Q}_0^{(1)}=\mathcal{Q}_0^{(2)}$. The coefficients $C^{(1)}$ and $C^{(2)}$ are
\begin{align*}
C^{(1)}=-\frac{2}{N}\left\lfloor{\frac{N_0+1}{2}}\right\rfloor\mathcal{Q}_0\hspace{.1cm}\delta_{\frac{M_0}{2},\lfloor{\frac{M_0}{2}}\rfloor}+\sum_{n=1}^{N_0}\sum_{q_0=1}^\frac{N-2}{2}A_{q_0}^{(n)}\hspace{.07cm}\mathcal{Q'}_{q_0}+\sum_{n\hspace{.06cm}\text{odd}}^{N_0}\sum_{q_0=1}^\frac{N-2}{2}B_{q_0}^{(N-1,n)}\hspace{.07cm}\mathcal{Q'}_{q_0},
\end{align*}
\begin{align*}
C^{(2)}=-\frac{2}{N}\left\lfloor{\frac{N_0+1}{2}}\right\rfloor\mathcal{Q}_0\hspace{.1cm}\delta_{\frac{M_0-1}{2},\lfloor{\frac{M_0}{2}}\rfloor}-\sum_{n=1}^{N_0}\sum_{q_0=1}^\frac{N-2}{2}A_{q_0}^{(n)}\hspace{.07cm}\mathcal{Q'}_{q_0}+\sum_{n\hspace{.06cm}\text{odd}}^{N_0}\sum_{q_0=1}^\frac{N-2}{2}B_{q_0}^{(1,n)}\hspace{.07cm}\mathcal{Q'}_{q_0}
\end{align*}
where $\mathcal{Q'}_{q_0}$ is defined similarly as before and
\begin{align*}
A_{q_0}^{(n)}=\frac{4}{N}(-1)^{n+1}&\sin^2\frac{\pi q_0n}{N}\\
&\sum_{p=0}^{\left\lfloor{\frac{M-3}{2}}\right\rfloor}\left[\sum_{k=0}^p\left(\begin{array}{c} M_0-1-k \\ k \end{array}\right)\left(\begin{array}{c} M-M_0-2-(p-k) \\ p-k \end{array}\right)\right](2\cos\frac{\pi q_0}{N})^{M-3-2p},
\end{align*}
\begin{align*}
B_{q_0}^{(N-1,n)}=&\frac{4}{N}(-1)^{q_0+1}(-1)^{\frac{n-1}{2}}\sin\frac{\pi q_0}{N}\sin\frac{\pi q_0n}{N}\\&\sum_{p=0}^{\left\lfloor{\frac{M-M_0-2+\delta_{\frac{M_0}{2},\lfloor{\frac{M_0}{2}}\rfloor}}{2}}\right\rfloor}\left( \begin{array}{c} M-M_0-2+\delta_{\frac{M_0}{2},\lfloor{\frac{M_0}{2}}\rfloor}-p \\ p \end{array} \right)(2\cos\frac{\pi q_0}{N})^{M-M_0-2+\delta_{\frac{M_0}{2},\lfloor{\frac{M_0}{2}}\rfloor}-2p},
\end{align*}
\begin{align*}
B_{q_0}^{(1,n)}=\frac{4}{N}(-1)^{\frac{n-1}{2}}\sin\frac{\pi q_0}{N}\sin\frac{\pi q_0n}{N}\sum_{p=0}^{\left\lfloor{\frac{M_0-\delta_{\frac{M_0}{2},\lfloor{\frac{M_0}{2}}\rfloor}}{2}}\right\rfloor}\left( \begin{array}{c} M_0-\delta_{\frac{M_0}{2},\lfloor{\frac{M_0}{2}}\rfloor}-p \\ p \end{array} \right)(2\cos\frac{\pi q_0}{N})^{M_0-\delta_{\frac{M_0}{2},\lfloor{\frac{M_0}{2}}\rfloor}-2p}.
\end{align*} 
The final result is briefly
\begin{align}
\mathbb{P}_\text{left-passage}(z)=-\frac{2}{N}\left\lfloor{\frac{N_0+1}{2}}\right\rfloor\hspace{.1cm}(\delta_{\frac{M_0}{2},\lfloor{\frac{M_0}{2}}\rfloor}+\delta_{\frac{M_0-1}{2},\lfloor{\frac{M_0}{2}}\rfloor})+\sum_{n\hspace{.06cm}\text{odd}}^{N_0}\sum_{q_0=1}^\frac{N-2}{2}(B_{q_0}^{(N-1,n)}+B_{q_0}^{(1,n)})\hspace{.07cm}\frac{\mathcal{Q'}_{q_0}}{\mathcal{Q}_0}
\label{leftpassagedis}
\end{align}

By appropriate arrangements of fugacities, we can compute generalizations of the left-passage probability through similar computations.

Again we are interested in the scaling limit of the above probability \eqref{leftpassagedis}. Following the approach in Sec. \ref{loopssurrounding}, we obtain
\begin{align}
\mathbb{P}_\text{left-passage}(z)\approx\frac{4}{N}\sum_{j}(-1)^{j+1}\frac{\sin(\frac{N_0\pi j}{N})}{\sin(\frac{2\pi j}{N})}\frac{\sinh(\frac{(M-M_0)\pi j}{N})}{\sinh(\frac{M\pi j}{N})}+\frac{4}{N}\sum_{j}\frac{\sin(\frac{N_0\pi j}{N})}{\sin(\frac{2\pi j}{N})}\frac{\sinh(\frac{M_0\pi j}{N})}{\sinh(\frac{M\pi j}{N})}-\frac{N_0}{N}
\label{leftpassageapprox}
\end{align}
where $j=\frac{N}{2}-q$ and we assume that $M_0$ is even. Based on the result of \cite{kenyondouble}, we expect \eqref{leftpassageapprox} to be the harmonic function of $z$ with boundary values 1 on the top and right sides of any rectangle $\{0<x<a, 0<y<b; \frac{a}{b}=\frac{M}{N}\}$ and zero elsewhere. It is
\begin{align}
\label{harmonic}
u(x,y)=&\sum_{n=1}^{\infty}\frac{-2}{n\pi}\hspace{.1cm}\frac{[1+(-1)^{n+1}]}{\sinh(\frac{n\pi b}{a})}\sin\frac{n\pi x}{a}\sinh\frac{n\pi y}{a}\nonumber\\
&+\sum_{n=1}^{\infty}\frac{-2}{n\pi}\hspace{.1cm}\frac{[1+(-1)^{n+1}]}{\sinh(\frac{n\pi a}{b})}\sinh\frac{n\pi x}{b}\sin\frac{n\pi y}{b}
\end{align}

In contrast to \eqref{harmonic}, there is no obvious symmetry with respect to $M\leftrightarrow N$ and $M_0\leftrightarrow N_0$ (equivalently $a\leftrightarrow b$ and $x\leftrightarrow y$) in \eqref{leftpassageapprox}. To recover such symmetry, we take advantage of complex analysis. If we keep the second term in \eqref{leftpassageapprox} and use the approximation $\sin(\frac{2\pi j}{N})\approx\frac{2\pi j}{N}$, the result \eqref{leftpassageapprox} turns out to be \eqref{harmonic} by considering the complex-valued functions $f_1(z)=\frac{\sin(\frac{N_0}{N}z)\sinh(\frac{M-M_0}{N}z)}{z\sin z\sinh(\frac{M}{N}z)}$ and $f_2(z)=\frac{\sin(\frac{N_0}{N}z)\sinh(\frac{M_0}{N}z)}{z\sin z\sinh(\frac{M}{N}z)}$ and using a result of Cauchy's residue theorem for $f(z)=f_1(z)+f_2(z)$,
\begin{align}
\text{pr.v.}\int_{-\infty}^{\infty}f(x)\dd x=2\pi i\sum\text{Res}f(z)+\pi i\sum\text{Res}f(z)
\label{residue}
\end{align}
where the first sum extends over all poles in the upper half-plane and the second over all simple poles on the real axis. 

We can easily see that the LHS of \eqref{residue} for $f(x)$ is zero. The RHS of \eqref{residue} reproduces six terms, due to the contribution of upper half-plane poles, nonzero simple poles on the real axis, and the simple pole at the origin for both $f_1(z)$ and $f_2(z)$. The first term in \eqref{leftpassageapprox} can be replaced by the minus of the term due to the contribution of simple poles of $f_1(z)$ on the real axis (except the pole at the origin). So, five terms replace the first term in \eqref{leftpassageapprox}, two of them due to the contribution of the simple pole at the origin for both $f_1(z)$ and $f_2(z)$ will cancel the third term in \eqref{leftpassageapprox}. The remaining three terms account for the contribution of upper half-plane ploes of $f_1(z)$, and all nonzero poles (in the upper half-plane and on the real axis) of $f_2(z)$, which together with the second term in \eqref{leftpassageapprox}, recover the RHS of \eqref{harmonic}.
This result is consistent with that of SLE$_4$ on a rectangular domain \cite{schrammleft}.

\vspace{.6cm}
{\bf Acknowledgments.} We are grateful to Vladimir Plechko for very helpful comments.
%============================================================================================%
\begin{appendices}
\section{Some useful formulas}
\label{useful}
Here we bring some formulas, referred in the paper. We have used the orthogonality relations
\begin{align}
\label{exp}
\frac{1}{M}\sum_{m=1}^{M}\exp\left[i\frac{2\pi(p+p'-1)m}{M}\right]=\delta_{p+p'-1,M}, \frac{1}{M}\sum_{m=1}^{M}(-1)^{m}\exp\left[i\frac{2\pi(p+p'-1)m}{M}\right]=\delta_{{p+p'-1},\frac{M}{2}}+\delta_{{p+p'-1},\frac{3M}{2}}
\end{align} 
and 
\begin{align}
\label{sin}
\frac{2}{N+1}\sum_{n=1}^{N} (-1)^{n+1}\sin\frac{\pi qn}{N+1}\sin\frac{\pi q'n}{N+1}=\delta_{q+q',N+1},\hspace{.8cm}   \sum_{n=1}^{N} (\pm1)^n\sin\frac{\pi qn}{N+1}\cos\frac{\pi q'n}{N+1}=0
\end{align}
in section \ref{grassmann} in the process of the evaluation of \eqref{grasspartition}, to factorize the partition function.

The summation formulas 
\begin{align}
\label{cos}
\sum_{k=1}^{n}\cos(kx)=\frac{1}{2}\left[-1+\frac{\sin(n+\frac{1}{2})x}{\sin\frac{1}{2}x}\right],
\end{align}
\begin{align}
\label{-cos}
\sum_{k=1}^{n}(-1)^k\cos(kx)=\frac{1}{2}\left[-1+\frac{(-1)^n\cos(n+\frac{1}{2})x}{\cos\frac{1}{2}x}\right]
\end{align}
and
\begin{align}
\label{cosh}
\sum_{k=1}^\infty\frac{\cos(kx)}{k^2+\alpha^2}=\frac{\pi}{2\alpha}\frac{\cosh\alpha(\pi-x)}{\sinh\alpha\pi}-\frac{1}{2\alpha^2}
\end{align}
valid for $0\leq x\leq2\pi$ \cite{gradstein}, have been used in section \ref{loopssurrounding} to approximate \ref{N12}.
We have also used the following products in section \ref{monomer},
\begin{align}
\prod_{p=1}^\frac{M-1}{2}4\left[\alpha^2+\cos^2\frac{\pi p}{M+1}\right]\equiv\frac{\left[\alpha+(1+\alpha^2)^\frac{1}{2}\right]^{M+1}-\left[\alpha-(1+\alpha^2)^\frac{1}{2}\right]^{M+1}}{4\alpha(1+\alpha^2)^{\frac{1}{2}}},
\label{prodlarge}
\end{align}
\begin{align}
\hspace{.9cm}\prod_{q=1}^{\frac{N-1}{2}}\left[2\cos(\frac{\pi q}{N+1})\right]=\sqrt{\frac{N+1}{2}}
\label{prodsmall}
\end{align}
valid for $M$ and $N$ odd \cite{chamberland}.
%============================================================================================%
\section{A combinatorial calculation}
\label{calculation}
Here we are going to justify how some combinatorial expressions such as $\left(\begin{array}{c} M-p \\ p \end{array}\right)$ and so, some relations to Chebyshev polynomials spring up all over the paper. We restrict ourselves to one example, the same argument applies to the others. 

The number of configurations such as the one in Figure \ref{fig1}, on a cylindrical strip of perimeter $M$, is equal to the number of solutions of the following equation:
\begin{align*}
y_1+...+y_{p+1}=M-1-p,\hspace{1.5cm}y_1,y_{p+1}\geq0,\ \ y_i\geq1,\hspace{.1cm} i=2,...,p
\end{align*}
which is equal to $\left(\begin{array}{c} M-p \\ p \end{array}\right)$, as well as solutions of the equation
\begin{align*}
y_1+...+y_p=M-p,\hspace{1.5cm} y_i\geq1,\hspace{.1cm} i=1,...,p
\end{align*}
where $p$ indecates the number of cross-forms in the cylindrical strip. The former gives the corresponding solution for a free strip, while the latter exclusively pertains to the additional solutions owing to the cylidrical condition. This accounts for the factor $\left[\left( \begin{array}{c} M-p \\ p \end{array} \right)+\left( \begin{array}{c} M-p-1 \\ p-1 \end{array} \right)\right]$ in \eqref{cylinderpartitioneven} and \eqref{cylinderpartitionodd}.
Now here is a connection between the partition function of the dimer model and Chebyshev polynomials of the second kind, $U_n(x)=\sin(n+1)\theta/\sin\theta,  \cos\theta=x$. By using the formula \cite{riordan}
\begin{align}
\sum_{k=0}^{\left\lfloor{\frac{n}{2}}\right\rfloor}\left( \begin{array}{c} n-k \\ k \end{array} \right)x^k=(-i)^nx^\frac{n}{2}U_n(\frac{i}{2\sqrt{x}})
\label{chebyshev}
\end{align}
the partition function \eqref{cylinderpartitioneven} can be rewritten as
\begin{align}
\mathcal{Q}_0^{cyl}=\prod_{q=1}^\frac{N}{2}\left[(-1)^\frac{M}{2}\left\{U_M(ix_q)-4x_q^2U_{M-2}(ix_q)\right\}+2\right]\hspace{.3cm}(N\hspace{.1cm}\text{even})
\label{chebyshevpartition}
\end{align}
and also \eqref{freepartitione} becomes
\begin{align}
\mathcal{Q}_0=\prod_{q=1}^\frac{N}{2}\left[(-1)^\frac{M}{2}U_M(ix_q)\right]\hspace{.3cm}(N\hspace{.1cm}\text{even})
\label{chebyshevpartitionf}
\end{align}
where $x_q=\cos\frac{\pi q}{N+1}$.

We can also use the following identities \cite{chamberland}
\begin{align}
&\prod_{p=1}^\frac{M}{2}2\left[\alpha^2+\sin^2\frac{\pi(2p-1)}{M}\right]^\frac{1}{2}\equiv\left[\alpha+(1+\alpha^2)^\frac{1}{2}\right]^\frac{M}{2}+\left[-\alpha+(1+\alpha^2)^\frac{1}{2}\right]^\frac{M}{2},\nonumber\\
&\prod_{p=1}^\frac{M}{2}4\left[\alpha^2+\cos^2\frac{\pi p}{M+1}\right]\equiv\frac{\left[\alpha+(1+\alpha^2)^\frac{1}{2}\right]^{M+1}-\left[\alpha-(1+\alpha^2)^\frac{1}{2}\right]^{M+1}}{2(1+\alpha^2)^\frac{1}{2}}
\label{eqkasteleyn}
\end{align}
valid for even $M$ and non-negative values of $\alpha$, and
\begin{align}
\sum_{k=0}^{\left\lfloor{\frac{n}{2}}\right\rfloor}\left( \begin{array}{c} n-k \\ k \end{array} \right)x^{n-k}=2^{-n-1}(x+4)^{-\frac{1}{2}}x^\frac{n}{2}\left[(x^\frac{1}{2}+(x+4)^\frac{1}{2})^{n+1}-(x^\frac{1}{2}-(x+4)^\frac{1}{2})^{n+1}\right]
\label{combinatorialidentity}
\end{align}
to check the results for the partition functions in section \ref{grassmann}.
%============================================================================================%
\section{A series formula}
\label{seriesformula}
 One brilliant consequence of contour integration, provided with some conditions, is the Poisson summatin formula
\begin{align}
\sum_{n\in\mathbb{Z}}f(n)=\sum_{n\in\mathbb{Z}}\hat{f}(n)
\label{poisson}
\end{align}
where $\hat{f}(\xi)=\int_{-\infty}^{\infty}f(x)e^{-2\pi ix\xi}\dd x$ is the Fourier transform of $f$. It is grounded in Cauchy's residue theorem and in turn has many far-reaching consequences in number theory, partial differential equations, statistical studies of time-series, and even improving the computability of slow convergent series \cite{complex, wiki, computability}. Several interesting and practically important identities have been derived from \eqref{poisson}, here we derive another one.
\vspace{.4cm}

{\bf Proposition.} For every $\alpha\in\mathbb{R}^*$ and $2m\in\mathbb{Z}^*$, the following identity holds
\begin{align}
\sum_{n=-\infty}^{\infty}\frac{1}{\cosh(\alpha n)+\cosh(\alpha m)}=\frac{2m}{\sinh(\alpha m)}.
\label{asli}
\end{align}
where by $\mathbb{R}^*$ and $\mathbb{Z}^*$ we mean $\mathbb{R}\setminus\{0\}$ and $\mathbb{Z}\setminus\{0\}$, respectively.

\vspace{.4cm}
An application of identity \eqref{asli} has appeared in section \ref{loopssurrounding}, where we have used the summation of odd and even terms separately. We believe this identity is of independent interest and could have general applications.

\begin{proof}
We consider the function $f(x)=\frac{1}{\cosh(\alpha z)+\cosh t}$, $t$ a non-zero parameter which is also real at the moment. First, we note that $f$ can be analytically continued in a horizontal strip containing the real axis, and there exists $A>0$ such that $|f(x)|\leq A/(1+x^2)$ for all $x\in \mathbb{R}$, i.e. $f$ has moderate decrease at infinity. These conditions are sufficient for the integral defining the Fourier transform to converge \cite{complex}.

We recall the calculation of Fourier transforms in \cite{complex}, which in this case obtains $\hat{f}(\xi)=\frac{2\pi}{\alpha \sinh t}\frac{\sin(\frac{2\pi t}{\alpha}\xi)}{\sinh(\frac{2\pi^2}{\alpha}\xi)}$. Applying \eqref{poisson} to $f$ and $\hat{f}$ leads to
\begin{align*}
\sum_{n=-\infty}^{\infty}\frac{1}{\cosh(\alpha n)+\cosh t}=\sum_{n=-\infty}^{\infty}\frac{2\pi}{\alpha \sinh t}\frac{\sin(\frac{2\pi t}{\alpha}n)}{\sinh(\frac{2\pi^2}{\alpha}n)}
\end{align*}
Now, if $2t/\alpha$ is a non-zero integer number, all the terms on the right-hand side will vanish with the exception of the one for $n=0$, which equals $\frac{2t/\alpha}{\sinh t}$. This results in the desired identity.
\end{proof}

It is interesting that if $t/\alpha\in\mathbb{Z}^*$, the odd and even terms of the series \eqref{asli} have equal contributions in summation. We can easily show that by, for example, the Fourier transform of $f(x)=\frac{1}{\cosh(2\alpha z)+\cosh t}$ for $t/\alpha\in\mathbb{Z}^*$.

\vspace{.6cm}
{\bf Remark.}  Though \eqref{asli} exclusively holds for $m\in\mathbb{Z}^*/2$, one can check that the RHS of \eqref{asli} is a very effective approximation for the aforementioned series if $|\alpha|$ is not so large (roughly, up to nearly 1 suffices).

%========================================================================================%
\end{appendices}
%========================================================================================%

%========================================================================================%
\end{document}